\documentclass[journal]{IEEEtran}
\usepackage{lipsum}
\usepackage[utf8]{inputenc}
\usepackage{amsmath,amsfonts,amssymb,bm}
\usepackage{graphicx}
\usepackage{subcaption}
\usepackage{xcolor}
\usepackage{mathtools}
\usepackage{comment}
\usepackage{textcomp}

\usepackage{multirow}
\usepackage{epstopdf}
\usepackage{flushend}
\usepackage{cite}
\usepackage{amssymb,relsize,graphicx,amsmath,psfig,cite,float,setspace,braket}
\usepackage{xcolor}
\usepackage{tabularx}
\usepackage{algorithm}
\usepackage{algpseudocode}
\usepackage{subcaption}
\usepackage[utf8]{inputenc}
\usepackage{amsfonts, bm}
\usepackage{caption}
\usepackage{comment}
\usepackage{textcomp}
\usepackage{booktabs,flushend}

\newcommand{\bi}{\begin{itemize}}
\newcommand{\ei}{\end{itemize}}

\newcommand{\be}{\begin{enumerate}}
\newcommand{\ee}{\end{enumerate}}
\newcommand{\bd}{\begin{description}}
\newcommand{\ed}{\end{description}}
\newcommand{\bc}{\begin{center}}
\newcommand{\ec}{\end{center}}
\newcommand{\bt}{\begin{tabbing}}
\newcommand{\et}{\end{tabbing}}
\newcommand{\bfig}{\begin{figure}}
\newcommand{\efig}{\end{figure}}
\newcommand{\beq}{\begin{equation}}
\newcommand{\beqarr}{\begin{eqnarray}}
\newcommand{\beqarrn}{\begin{eqnarray*}}
\newcommand{\eeq}{\end{equation}}
\newcommand{\eeqarr}{\end{eqnarray}}
\newcommand{\eeqarrn}{\end{eqnarray*}}
\newcommand{\bflr}{\begin{flushright}\vspace{-0.2in}}
\newcommand{\eflr}{\end{flushright}}
\newcommand{\bsub}{\begin{subequations}}
\newcommand{\esub}{\end{subequations}}
\newcommand{\barr}{\begin{array}}
\newcommand{\earr}{\end{array}}
\newcommand{\nn}{\nonumber}
\def\binom#1#2{\left( \!  \barr{c} #1 \\ #2 \earr \!  \right)}

\def\undb#1{\mbox{\bf{#1}}}

\usepackage{tikz}
\usetikzlibrary{arrows.meta,decorations.pathmorphing}

\definecolor{trustfill}{RGB}{232,240,250}
\definecolor{untrustfill}{RGB}{253,236,234}
\definecolor{untrustdraw}{RGB}{178,34,34}
\definecolor{titlered}{RGB}{140,0,0}
\definecolor{epsfill}{RGB}{255,247,218}
\definecolor{beamlight}{RGB}{160,192,222}
\definecolor{beamdark}{RGB}{100,145,198}
\definecolor{blob}{RGB}{200,215,232}
\definecolor{bgorange}{RGB}{220,140,0}
\definecolor{bgyellow}{RGB}{225,20,0}

\allowdisplaybreaks
\def\BibTeX{{\rm B\kern-.05em{\sc i\kern-.025em b}\kern-.08em
		T\kern-.1667em\lower.7exH_box{E}\kern-.125emX}}

\begin{document}

\title{Performance Analysis of Entanglement Swapping DV-QKD over MIMO FSO Channels}
\author{Sushil Kumar, Soumya~P.~Dash,~\IEEEmembership{Senior Member,~IEEE}, and George C. Alexandropoulos, \IEEEmembership{Senior Member, IEEE}
\thanks{S. Kumar and S.~P.~Dash are with the Department of Electronics and Communication Engineering, School of Electrical and Computer Sciences, Indian Institute of Technology Bhubaneswar, Argul, Khordha, 752050 India e-mail: (a24ec09010, spdash)@iitbbs.ac.in.}
\thanks{G. C. Alexandropoulos is with the Department of Informatics and Telecommunications, National and Kapodistrian University of Athens, Panepistimiopolis Ilissia, 16122 Athens, Greece, and also with the Department of Electrical and Computer Engineering, University of Illinois Chicago, IL 60601, USA (email: alexandg@di.uoa.gr).}
}
\maketitle

\begin{abstract}
This paper studies multiple-input multiple-output (MIMO) free-space optical (FSO) quantum communication systems employing polarization-encoded discrete-variable quantum key distribution (DV-QKD), based on entanglement swapping, with the goal of extending secure communications beyond the reach of a direct link. The FSO channels are subject to atmospheric absorption, diffraction, beam wander, pointing errors, and turbulence-induced fading, while background noise and detector imperfections degrade the performance. An eavesdropper controlling the channels and the relay is also considered. To analyze the performance of the considered system, both hops MIMO channels are decomposed into parallel spatial eigenchannels via singular value decomposition (SVD), photon loss is modeled using Kraus operators, and a threshold-detector model is developed for the relay. Based on these models, closed-form per-eigenchannel expressions for the relay gain are derived, which are then used to obtain expressions for the quantum bit error rate (QBER) and the secret key rate (SKR). Numerical results demonstrate that entanglement swapping maintains a low QBER over longer distances than direct transmission, extending the communication range beyond the noise-limited cutoff of the direct link, at the cost of a lower SKR at short distances. The crossover distance beyond which entanglement swapping prevails is determined mainly by the background noise and the detector technology, and it is demonstrated to increase with MIMO configurations. Moreover, the reach gain is shown to be maximized for a midpoint relay. Finally, since each swapping event requires a single-pair emission from both sources, a near-deterministic entangled-photon source is required, which spontaneous parametric down-conversion sources do not provide at practical key rates, whereas MIMO substantially relaxes this requirement.
\end{abstract}
\section{Introduction}
The upcoming sixth-generation (6G) networks aim to provide seamless connectivity anytime, anywhere~\cite{9144301,9145564}, which requires a dense terrestrial infrastructure to support large volumes of data flow between base stations, edge data centers, enterprise campuses, and critical facilities. Free-space optical (FSO) communications is a key enabling technology for such infrastructure, owing to its high capacity, immunity to radio-frequency interference, license-free operation, and rapid deployment over line-of-sight (LoS) links where laying new fiber is costly or slow~\cite{FSO_elamassie_2023_10111210, FSO6g_bekkali2023_10325456, 8920091}. However, since these terrestrial links interconnect sites of multiple operators within dynamic topologies, they require secure transmission in the presence of eavesdroppers who may compromise legitimate traffic.

Several studies have employed classical cryptographic techniques to secure data transmission in FSO systems~\cite{cryp_fso, 10411843, 6775279}, relying on computational complexity that conventional computers cannot overcome in practical time. However, Shor's algorithm efficiently breaks public-key schemes such as Rivest-Shamir-Adleman (RSA) and elliptic curve cryptography (ECC), and Grover's algorithm weakens symmetric-key schemes~\cite{shor1999polynomial, 11175376, Gitonga2025QuantumImpact}, so that traffic recorded today can be decrypted once large-scale quantum computers become available. Quantum key distribution (QKD) addresses this threat by exploiting fundamental principles of quantum mechanics, namely the no-cloning theorem and measurement-induced disturbance, enabling two legitimate parties to establish a secret key whose security is independent of the computational power of a potential eavesdropper~\cite{gisin2002quantum, 8732438, pirandola2020advances, pirandola2021limits}. Among QKD protocols, continuous-variable (CV) and discrete-variable (DV) schemes constitute the two main families~\cite{cv_dv_qkd_Scarani_2009, pirandola2020advances}. DV-QKD encodes the key onto single photons, typically in their polarization, which is only weakly affected by atmospheric propagation. It also benefits from mature security proofs and numerous experimental demonstrations~\cite{bennett1984update, pirandola2020advances}, including its entanglement-based form, in which the two users measure shared entangled pairs in randomly chosen bases and bound the information of an eavesdropper by the observed error rates~\cite{ekert1991quantum, bbm92, shor2000simple}. Owing to these advantages, the feasibility of polarization-based DV-QKD over free space has been demonstrated experimentally over terrestrial links of $144$~km in both prepare-and-measure and entanglement-based forms~\cite{schmittmanderbach2007decoy, ursin2007entanglement}, and in daylight environments over $53$~km~\cite{liao2017daylight}.

However, ground-level terrestrial FSO links suffer from several impairments, such as diffraction and atmospheric absorption, which attenuate the transmitted signal, as well as pointing errors, turbulence-induced beam wander, and fading, which cause random fluctuations of the received signal~\cite{Andrews2005, Vasylyevjun2012}. Consequently, background radiation and detector dark counts dominate the quantum bit error rate (QBER) as the signal weakens, and the secret key rate (SKR) vanishes beyond a noise-limited distance. Moreover, a long terrestrial link requires an unobstructed LoS, which is limited by buildings, terrain, and Earth's curvature in practical deployments. To mitigate turbulence-induced fading and increase the collected power, multiple-input multiple-output (MIMO) FSO systems employ arrays of transmit and receive apertures, providing spatial diversity and multiplexing gains~\cite{khalighi2014survey}. In the context of QKD, MIMO CV-QKD was studied over terahertz channels in~\cite{neel_spd_THZ_QKD_oct2021}, and this study was extended to account for channel estimation in~\cite{neel_spd_channel_estimation_skr}. Building on this, previous works exploited MIMO processing for QKD, analyzing CV-QKD MIMO terahertz channels enriched with a reconfigurable intelligent surface (RIS)~\cite{11555452, Sushil_RISCVQKD2025} and over MIMO FSO channels~\cite{sushil_11129674}, including hybrid quantum noise~\cite{sushil2026hybridnoise}, as well as DV-QKD with one- and two-way protocols over MIMO FSO channels~\cite{sushilicc2026}, and showed that MIMO improves the achievable SKR. Nevertheless, these works consider direct links between users, whose reach remains noise-limited and still requires an unobstructed LoS over the entire distance.

The latter limitations can be overcome by dividing the link into shorter segments via an intermediate relay, and entanglement swapping enables the same using an untrusted relay. In entanglement swapping~\cite{zukowski1993, pan1998}, Alice and Bob generate an entangled photon pair, retain one photon, and send the other to the relay, which performs a Bell-state measurement (BSM) on the two arriving photons. A successful outcome, announced publicly, projects the two retained photons onto an entangled state, even though they never interacted. As a result, each photon travels only over a shorter segment, and the relay accepts only coincident detections, which suppresses the impact of noise. Furthermore, since Alice and Bob measure their retained photons immediately and keep only the announced rounds, no quantum memory is required, and any misbehavior of the relay appears as errors in their correlations. Moreover, since users measure only photons from their own sources, their detectors are isolated from the channel, thereby making entanglement-swapping-based QKD the entanglement-based counterpart of measurement-device-independent (MDI) QKD~\cite{lo2012mdi}. However, these advantages come at a cost. A linear-optics BSM can identify only two of the four Bell states, which limits its success probability to $1/2$~\cite{calsamiglia2001}. Furthermore, each swapping event requires a pair from each source and the arrival of both photons at the relay, where the latter leads to a rate that scales with the product of the two segment transmissivities. Moreover, in the absence of quantum memories, the achievable rate remains bounded by the repeaterless limit~\cite{natural_comm2017pirandola}. Therefore, the benefit of entanglement swapping over terrestrial FSO channels depends on the trade-off between this penalty and the shorter, noise-filtered segments, which must be quantified under realistic channel and device conditions. 

In this direction, entanglement swapping with photon-number-encoded states was analyzed over lossy channels in~\cite{wan2024}, and entanglement-swapping-based QKD with practical sources and detectors was modeled over optical fiber in a single-input single-output (SISO) configuration in~\cite{Scherer2011practical}. Over atmospheric channels, Gaussian entanglement distribution and CV MDI-QKD were studied for satellite links~\cite{hosseinidehaj2018satellite,hosseinidehaj2017cvmdi}, while MDI-QKD has been demonstrated over an urban free-space channel~\cite{cao2020freespacemdi}, and analyzed over thermal-loss and phase-noise channels~\cite{peng2025mdi}. However, these studies consider fixed-loss fiber links, CV protocols, or satellite channels, and none of them addresses polarization-encoded entanglement swapping over terrestrial FSO channels with diffraction, pointing errors, turbulence, and practical threshold detectors. This gap is particularly relevant since every successful swapping event requires two photons to arrive over two independently fading channels; hence, the role of MIMO in such relay-based links cannot be inferred from direct-link studies. Consequently, the channel, noise, and device conditions under which the relay outperforms a direct link, the extent to which MIMO strengthens this advantage, and the source quality required to achieve it have not been identified.

Motivated by this research gap, in this paper, we study a terrestrial entanglement-swapping-based DV-QKD system supported by MIMO FSO links, in which Alice and Bob each hold a polarization-entangled photon-pair source and establish a secret key through an untrusted relay that performs a linear-optics BSM, while an eavesdropper controls both the channels and the relay. Since each swap event consumes one pair from each source, the sources' emission statistics play a decisive role. Practical sources based on spontaneous parametric down-conversion (SPDC) emit a random number of pairs per attempt~\cite{ma2007entangled}. Vacuum emissions, therefore, reduce the heralding rate, whereas multi-pair emissions lead to accepted BSM outcomes with uncorrelated measurement results~\cite{Scherer2011practical}. For this reason, we first develop the analysis in closed form for an ideal Bell-pair source, which provides an upper bound on the achievable performance, and then use the same framework to determine the source quality required by the relay and to assess whether SPDC sources satisfy it. The main contributions of this paper are summarized as follows:
\begin{itemize}
\item We develop a MIMO FSO channel model for the Alice-relay, Bob-relay, and direct Alice-Bob links that incorporates atmospheric absorption, diffraction, pointing errors, beam wander, log-normal fading in weak turbulence, correlated turbulence-induced path phases, and finite receiver apertures with single-mode coupling. We then apply singular value decomposition (SVD) on this channel to obtain the parallel eigenchannels on which the swapping attempts are performed.
\item We describe photon loss by Kraus operators and derive the state shared by Alice and Bob after a successful BSM on each paired eigenchannel, together with its success probability. Unlike~\cite{wan2024} which considers photon-number encoding over generic lossy channels, we treat polarization encoding over turbulent MIMO FSO channels and extend the loss model to multi-photon emissions.
\item We incorporate practical threshold detection, including dark counts, background clicks, the false heralds they cause at the relay, and imperfect two-photon interference, and derive closed-form per-eigenchannel expressions for the relay gain, the QBER in both bases, and the SKR, which are then averaged over the random MIMO channel.
\item By comparing the performance of our entanglement-swapping-based system with the system employing a direct link only under the same channel and detector models, we show that the relay extends the reach beyond the noise-limited cutoff of the direct link. We also quantify how the number of apertures, the background noise, the pointing jitter, the Hong-Ou-Mandel (HOM) visibility, and the detector technology determine the crossover distance beyond which the relay outperforms the direct link. We further show that the reach gain is maximized for a midpoint relay and remains positive over a broad range of relay positions.
\item We map the reach gain over the single- and multi-pair emission probabilities of a general source, and show that the advantage requires a near-deterministic entangled-photon source, which SPDC does not provide at practical key rates. Interestingly, the considered MIMO setup considerably relaxes this requirement.
\end{itemize}
The rest of the paper is organized as follows. Section~II describes the considered system model and the MIMO FSO channel model. Section~III presents the source model, the spatial eigenchannel decomposition, the Kraus description of photon loss, and the state shared by Alice and Bob after a successful BSM. Section~IV incorporates practical detection and derives the relay gain, QBER, and SKR of both the entanglement-swapping and direct links. Section~V presents the numerical results for the ideal Bell-pair source, the impact of the relay position, and the impact of a practical SPDC source. Finally, Section~VI concludes the paper.

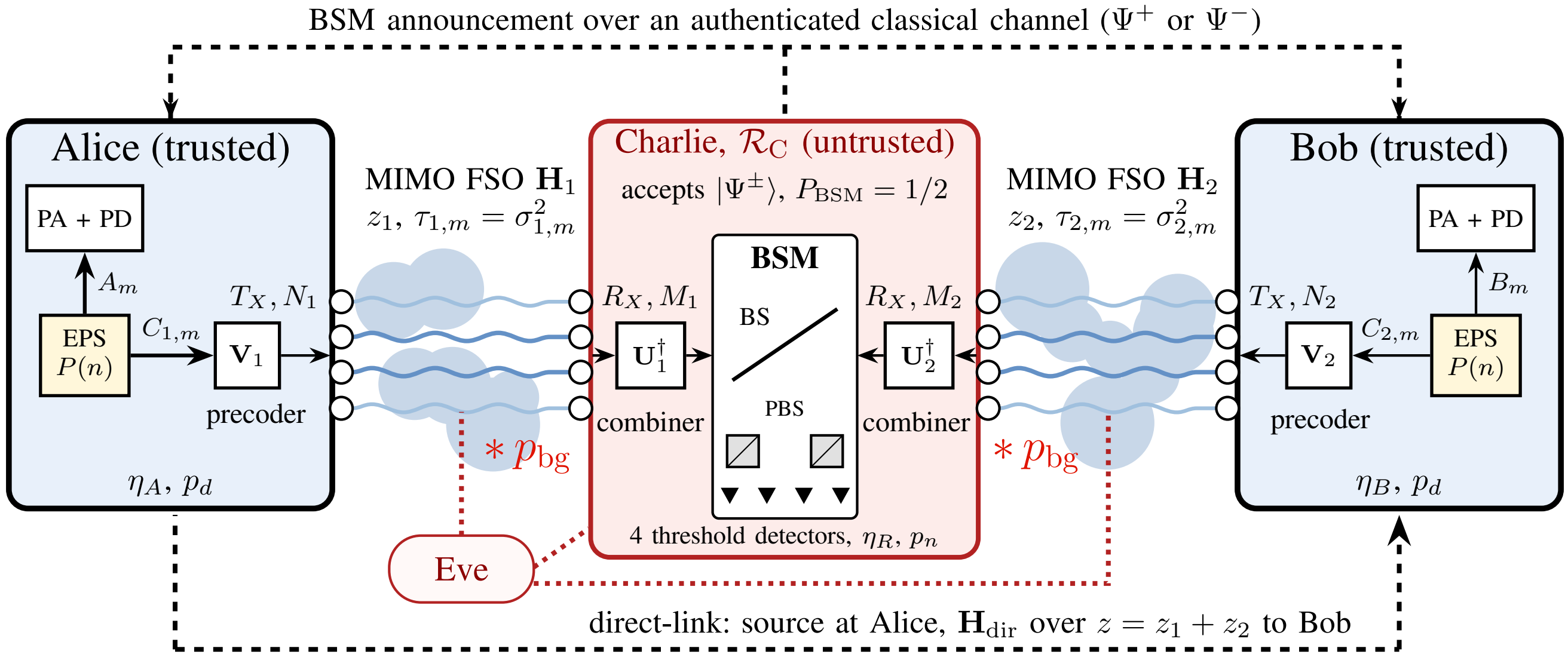
\begin{figure*}[t]
\centering
\centering
\begin{tikzpicture}[
  font=\footnotesize,
  >=Stealth,
  blk/.style={draw, line width=1pt, fill=white, minimum height=0.72cm, align=center},
  ap/.style={circle, draw, thick, fill=white, inner sep=0pt, minimum size=0.25cm},
  beamL/.style={line width=1.2pt, color=beamlight,
     decorate, decoration={snake, amplitude=0.45mm, segment length=7mm}},
  beamD/.style={line width=1.4pt, color=beamdark,
     decorate, decoration={snake, amplitude=0.45mm, segment length=7mm}},
  eveline/.style={densely dotted, thick, untrustdraw},
]

\begin{scope}[fill=blob, fill opacity=0.45]
  \fill (4.62,3.87) circle (0.43);  \fill (5.17,4.00) circle (0.45);
  \fill (4.85,2.95) circle (0.40);  \fill (5.27,2.93) circle (0.43);
  \fill (5.72,2.50) circle (0.52);  \fill (5.53,2.70) circle (0.30);
  \fill (5.93,2.80) circle (0.30);
  \fill (11.80,4.00) circle (0.52); \fill (12.13,3.54) circle (0.35);
  \fill (12.55,3.35) circle (0.30); \fill (13.30,3.75) circle (0.42);
  \fill (13.00,3.05) circle (0.32); \fill (12.52,2.52) circle (0.52);
  \fill (13.25,3.50) circle (0.45);
\end{scope}

\node[text=black,font=\fontsize{10pt}{11pt}\selectfont] at (8.95,6.98) {BSM announcement over an authenticated classical channel ($\Psi^{+}$ or $\Psi^{-}$)};
\draw[dashed, line width=1.5pt] (2.15,5.86) -- (2.15,6.68) -- (15.75,6.68) -- (15.75,5.86);
\draw[dashed, line width=1.5pt] (8.95,6.68) -- (8.95,5.86);
\draw[line width=1.5pt,thick,->] (2.15,6.10) -- (2.15,5.88);
\draw[line width=1.5pt,thick,->] (15.75,6.10) -- (15.75,5.88);

\draw[line width=1.8pt, rounded corners=6pt, fill=trustfill] (0.36,1.57) rectangle (3.94,5.86);
\node[text=black,font=\fontsize{13pt}{11pt}\selectfont] at (2.15,5.54) {Alice (trusted)};
\node[blk, minimum width=0.98cm] (paA) at (1.20,4.785) {PA + PD};
\node[blk, fill=epsfill, minimum width=0.98cm, minimum height=0.89cm] (epsA) at (1.20,3.265) {EPS\\$P(n)$};
\node[blk, minimum width=0.71cm] (V1) at (3.0,3.265) {$\mathbf{V}_1$};
\draw[line width=1.2pt,->] (epsA.north) -- node[right] {$A_m$} (paA.south);
\draw[line width=1.2pt,->] (epsA.east) -- node[above] {$C_{1,m}$} (V1.west);
\draw[thick,->] (V1.east) -- (3.92,3.265);
\node[text=black,font=\fontsize{9pt}{11pt}\selectfont] at (3.3, 3.90) {$T_X, N_1$};
\node[text=black,font=\fontsize{9pt}{11pt}\selectfont] at (3.1,2.59) {precoder};
\node[text=black,font=\fontsize{10pt}{11pt}\selectfont] at (2.15,1.82) {$\eta_A$, $p_d$};

\draw[line width=1.8pt, rounded corners=6pt, fill=trustfill] (13.96,1.57) rectangle (17.54,5.86);
\node[text=black,font=\fontsize{13pt}{11pt}\selectfont] at (15.75,5.54) {Bob (trusted)};
\node[blk, minimum width=0.98cm] (paB) at (16.60,4.785) {PA + PD};
\node[blk, fill=epsfill, minimum width=0.98cm, minimum height=0.89cm] (epsB) at (16.60,3.265) {EPS\\$P(n)$};
\node[blk, minimum width=0.71cm] (V2) at (14.86,3.265) {$\mathbf{V}_2$};
\draw[thick,->] (epsB.north) -- node[right] {$B_m$} (paB.south);
\draw[thick,->] (epsB.west) -- node[above] {$C_{2,m}$} (V2.east);
\draw[thick,->] (V2.west) -- (13.98,3.265);
\node[text=black,font=\fontsize{9pt}{11pt}\selectfont] at (14.58,3.90){$T_X, N_2$};
\node[text=black,font=\fontsize{9pt}{11pt}\selectfont] at (14.88,2.53) {precoder};

\node[text=black,font=\fontsize{10pt}{11pt}\selectfont] at (15.75,1.82) {$\eta_B$, $p_d$};

\draw[line width=1.5pt, draw=untrustdraw, rounded corners=6pt, fill=untrustfill] (6.80,1.03) rectangle (11.09,5.86);
\node[text=titlered, font=\fontsize{11pt}{11pt}\selectfont] at (8.95,5.61) {Charlie, $\mathcal{R}_{\rm C}$ (untrusted)};
\node[text=black,font=\fontsize{9pt}{11pt}\selectfont] at (8.95,5.1) {accepts $|\Psi^{\pm}\rangle$, $P_{\mathrm{BSM}}=1/2$};

\draw[thick, rounded corners=2pt, fill=white] (8.15,1.46) rectangle (9.75,4.60);
\node[font=\bfseries] at (8.95,4.35) {BSM};
\node at (8.62,3.70) {BS};
\draw[very thick] (8.37,2.99) -- (9.53,3.78);
\node[font=\scriptsize] at (8.95,2.68) {PBS};
\foreach \x in {8.31,9.24}{
  \draw[thick, fill=black!12] (\x,2.03) rectangle ++(0.35,0.35);
  \draw (\x,2.03) -- ++(0.35,0.35);
}
\foreach \x in {8.34,8.75,9.16,9.57}{
  \fill (\x-0.10,1.80) -- (\x+0.10,1.80) -- (\x,1.62) -- cycle;
}
\node at (8.95,1.25) {4 threshold detectors, $\eta_R$, $p_n$};

\node[blk, minimum width=0.75cm] (U1) at (7.465,3.265) {$\mathbf{U}_1^{\dagger}$};
\node[blk, minimum width=0.75cm] (U2) at (10.435,3.265) {$\mathbf{U}_2^{\dagger}$};
\draw[thick,->] (6.82,3.265) -- (U1.west);
\draw[thick,->] (U1.east) -- (8.15,3.265);
\draw[thick,->] (11.07,3.265) -- (U2.east);
\draw[thick,->] (U2.west) -- (9.75,3.265);

\node[text=black,font=\fontsize{9pt}{11pt}\selectfont] at (7.465,3.90) {$R_X, M_1$};
\node[text=black,font=\fontsize{9pt}{11pt}\selectfont] at (7.465,2.53) {combiner};

\node[text=black,font=\fontsize{9pt}{11pt}\selectfont] at (10.38,3.90) {$R_X, M_2$};
\node[text=black,font=\fontsize{9pt}{11pt}\selectfont] at (10.41,2.53) {combiner};
\foreach \y/\s in {3.86/beamL, 3.47/beamD, 3.08/beamD, 2.68/beamL}{
  \draw[\s] (4.16,\y) -- (6.57,\y);
  \draw[\s] (11.32,\y) -- (13.73,\y);
  \node[ap] at (4.04,\y) {};
  \node[ap] at (6.69,\y) {};
  \node[ap] at (11.20,\y) {};
  \node[ap] at (13.85,\y) {};
}
\node at (5.43,5.2) {\fontsize{10}{14} \selectfont MIMO FSO $\mathbf{H}_{1}$};
\node at (5.43,4.78) {\fontsize{10}{14} \selectfont $z_1$, $\tau_{1,m}=\sigma^2_{1,m}$};
\node[text=bgyellow] at (6.1,2.15){\fontsize{14}{14}\selectfont $*\, p_{\rm bg}$};
\node at (12.53,5.2) {\fontsize{10}{14} \selectfont MIMO FSO $\mathbf{H}_{2}$};
\node at (12.53,4.78) {\fontsize{10}{14} \selectfont $z_2$, $\tau_{2,m}=\sigma^2_{2,m}$};
\node[text=bgyellow] at (11.73,2.15){\fontsize{14}{14}\selectfont $*\, p_{\rm bg}$};

\draw[thick, draw=untrustdraw, rounded corners=10pt,
      fill=untrustfill!80!red!10]
      (4.57,0.53) rectangle (6.17,1.27);

\node[font=\fontsize{11pt}{11pt}\selectfont, text=titlered]
      at (5.37,0.90) {Eve};

\draw[draw=untrustdraw, line width=1.52pt, dotted]
      (5.37,1.27) -- (5.37,2.64);

\draw[draw=untrustdraw, line width=1.52pt, dotted]
      (6.17,0.92) -- (6.78,1.35);

\draw[draw=untrustdraw, line width=1.52pt, dotted]
      (6.17,0.74) -- (12.53,0.74) -- (12.53,2.64);
\node[text=black,font=\fontsize{10pt}{11pt}\selectfont] at (11,0.29) {direct-link: source at Alice, $\mathbf{H}_{\mathrm{dir}}$ over $z=z_1+z_2$ to Bob};
\draw[dashed, line width=1.5pt] (2.2,0.01) -- (15.75,0.01);
\draw[dashed, line width=1.5pt] (2.2,0.01) -- (2.2,1.5);
\draw[dashed, line width=1.5pt,->] (15.75,0.010) -- (15.75,1.5);
\end{tikzpicture}
\caption{The proposed MIMO FSO DV-QKD system between Alice and Bob via the entanglement swapping node $\mathcal{R}_{\rm C}$ in the presence of an eavesdropper Eve. EPS: Entangled photon source; PA: Polarization analyzer; PD: Photodetector; BS: Beam splitter; PBS: Polarizing beam splitter.}
\label{f1}
\vspace{-0.4cm}
\end{figure*}

\textit{Notations}: Boldface uppercase letters denote matrices, and $\mathbf{A}^\dagger$ denotes the conjugate transpose of a matrix $\mathbf{A}$. $\mathbf{I}_N$ represents the $N\times N$ identity matrix and ${\rm rank}(\cdot)$ denotes the rank of a matrix. $\|\cdot\|$ denotes the Euclidean norm and $\jmath\triangleq\sqrt{-1}$. $I_0(\cdot)$ denotes the zero-order modified Bessel function of the first kind. $\ket{\cdot}$ and $\bra{\cdot}$ denote a quantum state and its dual, $\otimes$ denotes
the tensor product, ${\rm Tr}(\cdot)$ denotes the trace, and ${\rm Tr}_{X}(\cdot)$ denotes the partial trace over subsystem $X$. Finally, $\mathbb{E}[\cdot]$ represents the expectation operator, where a subscript indicates the random variables over which the expectation is taken.
\section{System Model}
We consider a DV-QKD system based on entanglement swapping operating over MIMO FSO links, as shown in Fig.~\ref{f1}. Two legitimate nodes, Alice and Bob, wish to establish a secret key and are wirelessly connected through an untrusted relay, Charlie, denoted by $\mathcal{R}_{\rm C}$. To this end, there exist two FSO links between the legitimate nodes: link $k=1$ of distance $z_1$ joins Alice to $\mathcal{R}_{\rm C}$, and link $k=2$ of distance $z_2$ joins $\mathcal{R}_{\rm C}$ to Bob. The direct separation between Alice and Bob is denoted by $z$. Thus, $z_1+z_2=z$ if $\mathcal{R}_{\rm C}$ lies on the straight line joining Alice and Bob, and $z_1+z_2>z$ otherwise. Alice and Bob employ transmit arrays ($T_X$) of $N_1$- and $N_2$-element apertures, respectively. Each of them generates a pair of entangled photons and sends one photon to the intermediate node $\mathcal{R}_{\rm C}$. This relay node then employs two receive arrays ($R_X$) with $M_1$- and $M_2$-element  apertures, pointing toward Alice and Bob, respectively. The arriving photons from Alice and Bob are combined in a linear-optics Bell state measurement (BSM), and the BSM outcome is announced to them over an authenticated classical channel. Finally, as shown in Fig.~\ref{f1}, there exists an eavesdropper, Eve, that controls the loss of the transmitted photons in the FSO environment, restricting them from reaching Charlie. Following~\cite{shor2000simple}, the considered Eve employs the worst-case environmental access model.
\subsection{MIMO FSO Channel Model}
In the considered system model, the FSO channels between Alice and $\mathcal{R}_{\rm C}$, and Bob and $\mathcal{R}_{\rm C}$, are subjected to diffraction, atmospheric absorption, turbulence-induced irradiance and phase fluctuations, beam wander, and pointing errors. Denoting the corresponding complex MIMO FSO channel matrices by $\mathbf{H}_{1}\in\mathbb{C}^{M_1\times N_1}$ and $\mathbf{H}_{2}\in\mathbb C^{M_2\times N_2}$ (see Fig.~\ref{f1}), the coupling coefficient for $j$-th ($ j\in\{1,\ldots N_k\}$)  transmit aperture to the $i$-th ($ i\in\{1,\ldots M_k\}$) receive aperture in the $k$-th link is expressed as follows:
\beq
h_{ij}^{(k)}\triangleq\sqrt{T_{\mathrm{atm}}^{(k)} I_{i}^{(k)}} e^{\jmath \phi_{ij}^{(k)}}
\int_{\mathcal{D}_i} \! \! F_i^{*}(\mathbf{r})\, G_j^{(k)}\left(\mathbf{r},z_k;\tilde{\mathbf{r}}^{(k)}\right)
\mathrm{d}^2\mathbf{r}, \! \!
\label{eq1}
\eeq
where $T_{\mathrm{atm}}^{(k)} \triangleq 10^{-0.1\delta z_k}$ is the atmospheric absorption over link $k$ of length $z_k$ with $\delta$ (dB/km) being the wavelength-dependent atmospheric absorption coefficient, $I_{i}^{(k)}$ represents the irradiance fading at the $i$-th receiver, and $\phi_{ij}^{(k)}$ represents the turbulence-induced phase fluctuation of the path from the $j$-th transmitter to the $i$-th receiver. In addition, $\mathcal{D}_i$ denotes the circular pupil collection region of receiver $i$, given as:
\beq
\mathcal{D}_i \triangleq \left\{\mathbf{r}\in\mathbb{R}^2:
\|\mathbf{r}-\mathbf{b}_i\|\le a_r \right\},
\label{eq2}
\eeq
where $\mathbf{b}_i$ is the center and $a_r$ is the radius of $\mathcal{D}_i$.
In \eqref{eq1}, the integral describes geometrical and spatial-mode coupling, including diffraction, centroid displacement, finite-pupil collection, and receive-mode mismatch, where $G_j^{(k)}$ is the displaced and diffraction-limited field produced by the transmitter $j$ of the link $k$, which is given as follows:
\begin{align}
& \! \! G_j^{(k)} \left( \mathbf{r}, z_k; \tilde{\mathbf{r}}^{(k)}\right)
\triangleq \sqrt{\frac{2}{\pi w^2(z_k)}}
\exp \left( -\frac{ \|\mathbf{r}-\mathbf{a}_j-\tilde{\mathbf{r}}^{(k)}\|^2}{w^2(z_k)} \right) \nn \\
& \qquad \qquad \qquad
\times \exp \left( \frac{\jmath k_0 \|\mathbf{r}-\mathbf{a}_j-\tilde{\mathbf{r}}^{(k)}\|^2}
{2R(z_k)} - \jmath \zeta(z_k) \right) , \! \!
\label{eq3}
\end{align}
where $w(z_k)\triangleq w_0 \sqrt{1+\left(z_k/z_R\right)^2}$ represents the beam radius, with $w_0$ being the launched fundamental Gaussian mode at each transmitting aperture and $z_R\triangleq\frac{\pi w_0^2}{\lambda}$ the Rayleigh range. In addition, $k_0 \triangleq \frac{2\pi}{\lambda}$ is the wavenumber, $R(z_k) \triangleq z_k \left( 1 + \left(z_R/z_k \right)^2 \right)$ is the wavefront curvature, $\zeta(z_k) \triangleq \tan^{-1} (z_k/z_R)$ is the Gouy phase, $\mathbf{a}_j$ is the beam-center coordinate at the receive plane, and $\tilde{\mathbf{r}}^{(k)}$ represents random centroid displacement due to pointing error and beam wandering from turbulence. Its radial displacement $\|\tilde{\mathbf{r}}^{(k)}\|$ follows the Rayleigh distribution, with probability density function (p.d.f.):
\beq
f_{\|\tilde{\mathbf{r}}^{(k)}\|} \left(\tilde{\mathbf{r}}^{(k)}\right)
\triangleq \frac{\tilde{\mathbf{r}}^{(k)}}{\sigma_{\tilde{r},k}^2}
\exp\!\left(-\frac{\|\tilde{\mathbf{r}}^{(k)}\|^2}{2\sigma_{\tilde{r},k}^2} \right),
\quad \|\tilde{\mathbf{r}}^{(k)}\|\ge 0, 
\label{eq4}
\eeq
where $\sigma_{\tilde{r},k}^2$ is the per-axis variance of its two Cartesian components; it is obtained from the variances of the involved independent zero-mean Gaussian random variables as follows:
\beq
\sigma_{\tilde{r},k}^2 \triangleq
\sigma_{p,k}^2+\sigma_{\mathrm{BW},k}^2,
\label{eq5}
\eeq
where $\sigma_{p,k}^\triangleq z_k^2\theta_p^2$ is the pointing-error variance with $\theta_p$ denoting the per-axis pointing jitter angle, and $\sigma_{\mathrm{BW},k}^2 \triangleq 1.21\,C_n^2 z_k^{3} w_0^{-1/3}$ is the per-axis beam-wandering variance \cite{Andrews2005}, where $C_n^2$ is the refractive-index structure parameter. Under the Kolmogorov turbulence model~\cite{Andrews2005}, $C_n^2$ typically ranges from $10^{-17}\,\mathrm{m}^{-2/3}$ for weak turbulence to $10^{-14}\,\mathrm{m}^{-2/3}$ for moderate turbulence.

Moreover, in \eqref{eq1}, $F_i(\undb{r})$ represents the spatial mode profile of the $i$-th receiver's detection optics. At the receiver plane, the spatial coupling of the optical field is modeled via projection onto a finite set of detection modes defined by the physical receiving optics. For the $i$-th receiver element centered at the transverse coordinate $\mathbf{b}_i$ with a physical pupil radius $a_r$, the pupil-truncated spatial mode is defined as follows:
\beq
\!\!\!F_i\left(\mathbf{r}\right) \triangleq \xi_{\mathrm{rx}}
\exp \left( \! - \frac{\|\mathbf{r}-\mathbf{b}_i\|^2} {w_{\mathrm{rx}}^2}
\!+\!\frac{\jmath k_0\|\mathbf{r}-\mathbf{b}_i\|^2}{2R\left(z_k\right)} \right) \! ,
\mathbf{r}\in \mathcal{D}_i,
\label{eq6}
\eeq
where $w_{\mathrm{rx}}$ is the acceptance-mode radius determined by the receiver optics, and $\xi_{\mathrm{rx}}$ is the normalization constant that ensures $\int_{\mathcal{D}_i}|F_i(\mathbf{r})|^2\,\mathrm{d}^2\mathbf{r}=1$ and is given as:
\beq
\xi_{\mathrm{rx}} \triangleq
\sqrt{ 2 \left( \pi w_{\mathrm{rx}}^2
\left(1-\exp\left(-2a_r^2/w_{\mathrm{rx}}^2\right) \right) \right)^{-1}} \, .
\label{eq7}
\eeq

Experimental studies have shown that turbulence-induced fading in long-distance FSO communication channels can be effectively modeled using a lognormal distribution~\cite{Capraro_turbulence_2012}. Accordingly, the p.d.f. of the normalized irradiance fading coefficient $I_{i}^{(k)}$ for the $i$-th receiver is given as follows:
\beq
f_{I_{i}}\left(u\right) \triangleq
\frac{1}{u\sqrt{2\pi\sigma_{i}^2}}
\exp \left(-\frac{\left(\ln u+0.5\sigma_{i}^2\right)^2}{2\sigma_{i}^2} \right) , \ u>0,
\label{eq8}
\eeq
where $\sigma_{i}^2\triangleq\ln\left(1+\sigma_{I,i}^2\right)$ denotes the log-irradiance variance with $\sigma_{I,i}^2$ being the scintillation index. Since the beam is displaced, the $i$-th receiver lies at the radial offset $r_i=\|\mathbf{b}_i-\tilde{\mathbf{r}}^{(k)}\|$ from the beam center, and under weak turbulence, $\sigma_{I,i}^2$ is evaluated for the collimated Gaussian beam as~\cite{Andrews2005}:
\begin{align}
\sigma_{I,i}^2=& A_k \Big(3.86\,\chi_k^2\Big(0.40\left(\left(1+2\Theta_k\right)^2+4\Lambda_k^2\right)^{5/12} \nn \\
& \left. \times \cos \left( \frac{5}{6}\tan^{-1}\!\left(\frac{1+2\Theta_k}{2\Lambda_k} \right) \right)
-\frac{11}{16}\Lambda_k^{5/6} \right) \nn\\
& \left. + \, 4.42\,\chi_k^2\Lambda_k^{5/6}\frac{r_i^2}{w^2\left(z_k\right)} \right) ,
\label{eq9}
\end{align}
where $\chi_k^2\triangleq1.23\,C_n^2\,k_0^{7/6}z_k^{11/6}$ denotes the Rytov variance associated with the $k$-th propagation link, $\Lambda_k\triangleq\frac{z_k/z_R}{1+(z_k/z_R)^2}$, and $\Theta_k\triangleq\frac{1}{1+(z_k/z_R)^2}$ are the Gaussian-beam parameters at the receiver, and $A_k\triangleq\left(1+0.333\,k_0 w_{\mathrm{rx}}^2/z_k\right)^{-7/6}$ represents the aperture-averaging factor. The turbulence-induced phase fluctuation, $\phi_{ij}^{(k)}$, in~\eqref{eq1} is random and is accumulated along the path from the $j$-th transmitter to the $i$-th receiver, given as
\beq
\phi_{ij}^{(k)}=k_0\int_0^{z_k} n_1\!\left(\mathbf{a}_j+\frac{z}{z_k}\left(\mathbf{b}_i-\mathbf{a}_j\right),z\right)\mathrm{d}z,
\label{eq10}
\eeq
where $n_1(\cdot,\cdot)$ represents the refractive-index fluctuation of the atmosphere. Since $\phi_{ij}^{(k)}$ results from many independent refractive-index contributions along the path, it is a zero-mean Gaussian random variable with variance $\sigma_\phi^2$, which is dominated by the largest eddies and depends on the outer scale of turbulence. The mean-square phase difference between the paths $(i,j)$ and $(i',j')$ is obtained from~\eqref{eq10} under the Kolmogorov spectrum~\cite{Andrews2005}, as follows:
\begin{align}
D_{\phi}(i,j;i',j')
&= 2.91\,k_0^2 C_n^2 z_k \nn\\
&\hspace{-1.42cm}\times \int_0^1
\left\|
(1-u)(\mathbf{a}_j-\mathbf{a}_{j'})
+u(\mathbf{b}_i-\mathbf{b}_{i'})
\right\|^{5/3}\,\mathrm{d}u,
\label{eq11}
\end{align}
where $u\triangleq z/z_k$ represents the normalized propagation distance. Note that~\eqref{eq11} assumes that the outer scale is much larger than the array. Accordingly, the phases $\phi_{ij}^{(k)}$'s are jointly Gaussian with covariance $\mathbb{E}\big[\phi_{ij}^{(k)}\phi_{i'j'}^{(k)}\big]=\sigma_\phi^2-D_{\phi}(i,j;i',j')/2$. The coupling integral in~\eqref{eq1} is thus evaluated by substituting~\eqref{eq3} and~\eqref{eq6} into~\eqref{eq1}, followed by algebraic simplifications, yielding using $s_{ij}\triangleq\|\mathbf{b}_i-\mathbf{a}_j-\tilde{\mathbf{r}}^{(k)}\|$:
\begin{align}
h_{ij}^{(k)} & = \frac{4\sqrt{T_{\mathrm{atm}}^{(k)}I_{i}^{(k)}}
e^{\jmath \left(\phi_{ij}^{(k)}+\frac{k_0s_{ij}^2}{2R(z_k)}-\zeta\left(z_k\right)\right)}}
{w_{\mathrm{rx}} w\left(z_k\right)
\sqrt{1-\exp \left(-2a_r^2/w_{\mathrm{rx}}^2\right)}} \label{eq12}  \\
& \times \int_{0}^{a_r} \! \! \! \rho
e^{-\frac{\rho^2}{w_{\mathrm{rx}}^2}-\frac{\rho^2+s_{ij}^2}{w^2(z_k)}}
I_0 \left( \rho s_{ij} \left( \frac{\jmath k_0}{R(z_k)}-\frac{2}{w^2(z_k)} \right) \! \right) \mathrm{d} \rho.\nonumber
\end{align}
\section{Entanglement Swapping over MIMO FSO Links}
This section describes the entanglement-swapping process over the MIMO FSO links under pure-loss propagation. We specify the source states, decompose the MIMO channels into parallel spatial channels, and describe the interaction of the traveling photons with the loss environment using Kraus operators. The resulting states are then used to derive the state shared by the legitimate nodes Alice and Bob after a successful BSM at $\mathcal{R}_{\rm C}$ and the probability of this event.
\subsection{Source Model and Initial States}
In the entanglement-swapping protocol, Alice and Bob independently generate polarization-entangled photon pairs. Alice's source generates the pair $(A,C_1)$, of which she retains photon $A$ and transmits photon $C_1$ to $\mathcal{R}_{\rm C}$ via $\mathbf{H}^{(1)}$. Similarly, Bob's source generates the pair $(C_2,B)$, of which he retains photon $B$ and transmits photon $C_2$ to $\mathcal{R}_{\rm C}$ via $\mathbf{H}^{(2)}$. The qubit is encoded in the polarization of the photons. Since a practical source does not emit exactly one pair per attempt, its output state is modeled in the photon-number basis as follows:
\beq
\hat{\rho}_{\rm s} \triangleq \sum_{n=0}^{\infty} P\left(n\right)\,\ket{\Phi_n}\bra{\Phi_n},
\label{eq13}
\eeq
where $P(n)$ represents the probability that the source emits exactly $n$ pairs per attempt, and $\ket{\Phi_n}$ represents the normalized $n$-pair polarization-entangled state \cite{ma_jul2005}, which is given as:
\beq
\ket{\Phi_n} \triangleq \frac{1}{\sqrt{n+1}} \sum_{q=0}^{n}
\ket{n-q,\,q}_{a}\,\ket{n-q,\,q}_{b},
\label{eq14}
\eeq
where $\ket{i,j}_{a}$ and $\ket{i,j}_{b}$ represent $i$ horizontally and $j$ vertically polarized photons in the retained mode $a$ and the transmitted mode $b$, respectively, and $q$ represents the number of pairs emitted in the $VV$ configuration. Accordingly, the ideal Bell-pair and SPDC sources (considered in this work) differ in their pair-number distributions, in particular:
\beq
P(n) =
\begin{cases}
\delta_{n,1}, & \text{ideal Bell-pair source}\\
\dfrac{\left(n+1\right) \left(\mu/2\right)^{n}}{\left(1+\mu/2\right)^{n+2}}, & \text{SPDC source}
\end{cases},
\label{eq15}
\eeq
where $\delta_{n,1}$ represents the Kronecker delta and $\mu$ denotes the mean number of pairs emitted per attempt by the SPDC source, whose pair-number distribution is given as follows:
\beq
P\left(0\right)=1-p_1-p_2, \,\, P\left(1\right)=p_1,\, \, P\left(2\right)=p_2=\frac{g\,p_1^{2}}{2},
\label{eq16}
\eeq
Here, $p_1$ and $p_2$ represent the single-pair and two-pair emission probabilities, respectively, and $g\triangleq 2p_2/p_1^{2}$ represents the multi-pair parameter. Substituting~\eqref{eq16} into the definition of the second-order correlation function of the emitted pair number, $g^{\left(2\right)}\left(0\right)=\sum_n n\left(n-1\right)P\left(n\right)/\big(\sum_n nP\left(n\right)\big)^{2}$~\cite{Loudon2000}, yields $g^{\left(2\right)}\left(0\right)={g}/{\left(1+g\,p_1\right)^{2}}$, so that $g\simeq g^{\left(2\right)}\left(0\right)$ for $gp_1\ll1$. In this representation, the ideal Bell-pair source corresponds to $p_1=1$ and $g=0$, whereas the SPDC source corresponds to $g\simeq1.5$ for $\mu\ll1$. For the ideal Bell-pair source, i.e., $P(n)=\delta_{n,1}$, \eqref{eq14} reduces to the Bell state $\ket{\Phi^{+}}$, and the two initially generated states are given as
\begin{align}
\ket{\Phi^{+}}_{AC_{1}}
& \triangleq \frac{1}{\sqrt{2}}
\left(\ket{HH}_{AC_{1}} + \ket{VV}_{AC_{1}} \right),
\nn \\
\ket{\Phi^{+}}_{C_{2}B}
& \triangleq \frac{1}{\sqrt{2}}
\left( \ket{HH}_{C_{2}B} + \ket{VV}_{C_{2}B} \right),
\label{eq17}
\end{align}
where $\ket{H} \triangleq \ket{1,0}$ and $\ket{V} \triangleq \ket{0,1}$ represent a single photon with horizontal and vertical polarization, respectively. These two orthogonal states form the $Z$ basis, in which the bit values are encoded as $\ket{0_Z} = \ket{H}$ and $\ket{1_Z} = \ket{V}$. The density operator associated with either Bell pair is given as follows:
\begin{align}
   \rho^{\Phi^+} \triangleq \ket{\Phi^+}\bra{\Phi^+}
&= \frac{1}{2}\left(\ket{HH}\bra{HH} + \ket{HH}\bra{VV} \right. \nn \\
& \left.\quad+ \ket{VV}\bra{HH} + \ket{VV}\bra{VV}\right).
\label{eq18}
\end{align}
Since the two sources operate independently, the joint initial state is given as $\rho_{\rm in} = \rho^{\Phi^+}_{AC_{1}} \otimes \rho^{\Phi^+}_{C_{2}B}$.
\subsection{Spatial Eigenchannels Processing and their Pairing}
The transmitted photons $C_1$ and $C_2$ propagate to $\mathcal{R}_{\rm C}$ over links $k=1$ and $k=2$, respectively. As described in Section~II.A, the channel matrix $\mathbf{H}_{k}$ of each $k$-th link couples every transmit aperture to every receive aperture, so that a photon launched from a single aperture would be distributed over several receive modes. To obtain independent spatial channels over which the swapping protocols can be carried out in parallel, $\mathbf{H}_k$ is decomposed by the singular value decomposition (SVD) as follows:
\beq
\mathbf{H}_k = \mathbf{U}_k\boldsymbol\Sigma_k\mathbf{V}_k^\dagger \, ,
\label{eq19}
\eeq
where $\mathbf{U}_k\in\mathbb{C}^{M_k\times M_k}$ and $\mathbf{V}_k\in\mathbb{C}^{N_k\times N_k}$ represent unitary matrices, and $\boldsymbol\Sigma_k\in\mathbb{R}^{M_k\times N_k}$ contains the singular values $\sigma_{k,1}\ge\cdots\ge\sigma_{k,r_k}>0$, where $r_k=\operatorname{rank} \left(\mathbf{H}^{(k)}\right) \le \min\left(M_k,N_k\right)$. Considering the availability of perfect channel state information \cite{pang2020experimental} and linear-optical processing\cite{calsamiglia2001, braunstein1995measurement} networks that are reconfigured within the channel coherence time, Alice and Bob apply the precoding unitaries $\mathbf{V}_1$ and $\mathbf{V}_2$, respectively, whereas $\mathcal{R}_{\rm C}$ applies the combining unitaries $\mathbf{U}_1^\dagger$ and $\mathbf{U}_2^\dagger$ to the modes received from Alice and Bob, respectively. The same spatial transformations are applied to both polarization components, thereby preserving the polarization-encoded states. Accordingly, each link is diagonalized as follows:
\beq
\mathbf{U}_k^\dagger\mathbf{H}_k\mathbf{V}_k = \boldsymbol\Sigma_k,
\label{eq20}
\eeq
which separates it into $r_k$ parallel spatial eigenchannels, whose transmissivities are given $\forall m=1,\ldots,r_k$ as:
\beq
\tau_{k,m} = \sigma_{k,m}^2.
\label{eq21}
\eeq

Since the MIMO FSO channel does not provide any optical gain, all singular values of $\mathbf{H}_{k}$ satisfy $\sigma_{k,m} \le 1$, so that $0 \le \tau_{k,m} \le 1$. To carry out BSMs simultaneously over all the MIMO FSO channels, the eigenchannels of the two links are paired in decreasing order of their transmissivities, i.e., the $m$-th eigenchannel of link $k=1$ is paired with the $m$-th eigenchannel of link $k=2$, where $m=1,\ldots,r_{\rm ES}$ with $r_{\rm ES}\triangleq\min(r_1,r_2)$ representing the total number of available eigenchannel pairs. Accordingly, the photons involved in the $m$-th attempt are hereafter denoted by $A_m$, $C_{1,m}$, $C_{2,m}$, and $B_m$, where the traveling photons $C_{1,m}$ and $C_{2,m}$ experience independent pure-loss channels with transmissivities $\tau_{1,m}$ and $\tau_{2,m}$, respectively. The effect of these channels on the polarization-entangled states is described next.
\subsection{Kraus Evolution of the Entangled States}
On each $m$-th paired eigenchannel, the traveling photon $C_{k,m}$ is either transmitted to $\mathcal{R}_{\rm C}$ or lost to the environment; the FSO environment is inaccessible to Alice, Bob, and $\mathcal{R}_{\rm C}$ and is conservatively attributed to Eve. Accordingly, the state shared by each legitimate node and $\mathcal{R}_{\rm C}$ before the BSM is obtained by tracing out the environment, which can be equivalently expressed as a completely positive trace-preserving (CPTP) map in Kraus operator-sum form acting only on $C_{k,m}$. This map is first derived for the ideal Bell-pair source in~\eqref{eq17}, for which each transmitted mode carries a single photon, and then extended to multi-photon emissions.
Since photon loss maps the polarization qubit $\{\ket{H},\ket{V}\}$ onto the vacuum state $\ket{\mathrm{vac}}=\ket{0_H,0_V}$, the Hilbert space of $C_{k,m}$ is extended to the basis $\{\ket{H},\ket{V},\ket{\mathrm{vac}}\}$. Assuming polarization-independent propagation and spatial processing, both polarization components experience the same transmissivity $\tau_{k,m}$. The traveling photon $C_{k,m}$ interacts with an environmental mode $E_{k,m}$, initially in the vacuum state, and the corresponding Stinespring dilation is given by the following expression with $\alpha\in\{H,V\}$:
\begin{align}
&\mathbf{U}_{C_{k,m}E_{k,m}}
\ket{\alpha}_{C_{k,m}}\ket{\mathrm{vac}}_{E_{k,m}} \triangleq
\sqrt{\tau_{k,m}}\, \ket{\alpha}_{C_{k,m}}\ket{\mathrm{vac}}_{E_{k,m}}
\nn \\
&\quad +\sqrt{1-\tau_{k,m}}\,\ket{\mathrm{vac}}_{C_{k,m}}\ket{\alpha}_{E_{k,m}},
\label{eq22}
\end{align}
whereas the joint vacuum state remains invariant, i.e.,
\beq
\mathbf{U}_{C_{k,m}E_{k,m}}
\ket{\mathrm{vac}}_{C_{k,m}}\ket{\mathrm{vac}}_{E_{k,m}} \! =
\ket{\mathrm{vac}}_{C_{k,m}}\ket{\mathrm{vac}}_{E_{k,m}}.
\label{eq23}
\eeq
In the latter expressions, $E_{k,m}$ represents the inaccessible optical degrees of freedom into which the photon is lost, and the orthogonal environment states $\ket{H}_{E_{k,m}}$ and $\ket{V}_{E_{k,m}}$ retain the polarization of the lost photon. Projecting the output environment onto the basis $\{\ket{\mathrm{vac}}_{E_{k,m}},\ket{H}_{E_{k,m}}, \ket{V}_{E_{k,m}}\}$, yields the following Kraus operators:
\beq
K_{n,m}^{(k)} \triangleq
{}_{E_{k,m}}\!\bra{e_n}\, \mathbf{U}_{C_{k,m}E_{k,m}}\, \ket{\mathrm{vac}}_{E_{k,m}},
\label{eq24}
\eeq
where $\ket{e_n}\in\{\ket{\mathrm{vac}},\ket{H},\ket{V}\}$. On the ordered basis $\{\ket{H},\ket{V},\ket{\mathrm{vac}}\}$, the Kraus operators are given as follows:
\begin{subequations}
\begin{align}
K_{0,m}^{(k)}&= \sqrt{\tau_{k,m}}\left(\ket{H}\bra{H}+\ket{V}\bra{V}\right)
+\ket{\mathrm{vac}}\bra{\mathrm{vac}} \nn \\
&= \begin{pmatrix}
\sqrt{\tau_{k,m}} & 0 & 0\\
0 & \sqrt{\tau_{k,m}} & 0\\
0 & 0 & 1
\end{pmatrix},
\label{eq25a}
\\
&\hspace{-0.81cm} K_{1,m}^{(k)}=
\sqrt{1-\tau_{k,m}}\ket{\mathrm{vac}}\bra{H}=\!\!\begin{pmatrix}
0 & 0 & 0\\
0 & 0 & 0\\
\sqrt{1-\tau_{k,m}} & 0 & 0
\end{pmatrix} \!\!,
\label{eq25b}
\\
&\hspace{-0.81cm}K_{2,m}^{(k)}=
\sqrt{1-\tau_{k,m}}\ket{\mathrm{vac}}\bra{V} = \!\!\begin{pmatrix}
0 & 0 & 0\\
0 & 0 & 0\\
0 & \sqrt{1-\tau_{k,m}} & 0
\end{pmatrix} \! ,
\label{eq25c}
\end{align}
\end{subequations}
with $K_{0,m}^{(k)}$ representing the no-loss event, whereas $K_{1,m}^{(k)}$ and $K_{2,m}^{(k)}$ denote the losses of an $H$- and a $V$-polarized photon, respectively, which map the corresponding state onto the vacuum. The reduced pure-loss channel mapping is trace-preserving by satisfying the following relation:
\beq
\sum_{n=0}^{2} \left(K_{n,m}^{(k)}\right)^\dagger K_{n,m}^{(k)} = \mathbf{I}_3.
\label{eq26}
\eeq

On link $k=1$, the channel acts only on the traveling photon $C_{1,m}$, whereas Alice's retained photon $A_m$ is unaffected. Hence, the reduced state of the Alice-$\mathcal{R}_{\rm C}$ system is given as:
\begin{align}
\rho_{A_m C_{1,m}}^{\Phi^{+}} &= \sum_{n=0}^{2}
\left(\mathbf{I}_{A_m}\otimes K_{n,m}^{(1)} \right) \rho_{\mathrm{in}}^{A_m C_{1,m}}
\left( \mathbf{I}_{A_m}\otimes K_{n,m}^{(1)} \right)^\dagger
\nn \\
&=\frac{\tau_{1,m}}{2}
\left(\ket{HH}\bra{HH} + \ket{HH}\bra{VV}\right)_{A_m C_{1,m}}
\nn \\
+& \frac{\tau_{1,m}}{2}\left(\ket{VV}\bra{HH} + \ket{VV}\bra{VV}\right)_{A_m C_{1,m}}
 \label{eq27} \\
&\hspace{-1.50cm} +\frac{1-\tau_{1,m}}{2}
\left( \ket{H,\mathrm{vac}}\bra{H,\mathrm{vac}}
+ \ket{V,\mathrm{vac}}\bra{V,\mathrm{vac}} \right)_{A_m C_{1,m}}\!.
\nn
\end{align}
Similarly, on link $k=2$, the channel acts only on $C_{2,m}$, whereas Bob's retained photon $B_m$ is unaffected, so that the reduced state of the Bob-$\mathcal{R}_{\rm C}$ system is given by:
\begin{align}
\rho_{C_{2,m}B_m}^{\Phi^{+}} &=
\sum_{n=0}^{2} \left(K_{n,m}^{(2)}\otimes \mathbf{I}_{B_m}\right)
\rho_{\mathrm{in}}^{C_{2,m}B_m} \left(K_{n,m}^{(2)}\otimes \mathbf{I}_{B_m}\right)^\dagger
\nn \\
& = \frac{\tau_{2,m}}{2} \left(\ket{HH}\bra{HH} + \ket{HH}\bra{VV}\right)_{C_{2,m}B_m}
\nn \\
&+ \frac{\tau_{2,m}}{2}\left(\ket{VV}\bra{HH} + \ket{VV}\bra{VV} \right)_{C_{2,m}B_m}
\label{eq28} \\
&\hspace{-1.6cm}+\!\frac{1-\tau_{2,m}}{2} \left(\ket{\mathrm{vac},H} \!\bra{\mathrm{vac},H}
\!+\!\ket{\mathrm{vac},V}\!\bra{\mathrm{vac},V}\right)_{C_{2,m}B_m}\!.
\nn  
\end{align}
In the latter two expressions, the terms proportional to $\tau_{k,m}$ preserve the entangled state and correspond to successful transmission, whereas the terms proportional to $1-\tau_{k,m}$ correspond to erasure events in which the relay receives vacuum, and the correlation with the retained photon is lost.
For the SPDC and general sources in~\eqref{eq15} and~\eqref{eq16}, each polarization mode of $C_{k,m}$ may carry more than one photon. In this case, the pure-loss channel acts independently on every photon, and the Kraus operators in~\eqref{eq25a}-\eqref{eq25c} are generalized for each polarization mode to the following expression:
\beq
\hat{K}_{l,m}^{(k)} = \sum_{n=l}^{\infty}
\sqrt{\binom{n}{l}}\,\tau_{k,m}^{(n-l)/2}\,(1-\tau_{k,m})^{l/2}\,
\ket{n-l}\bra{n}, \! \!
\label{eq29}
\eeq
where $n$ and $l$ represent the number of photons in the polarization mode before propagation and lost to the environment, respectively. Accordingly, the operator acting on $C_{k,m}$ is given by $\hat{K}_{l_H,m}^{(k)}\otimes\hat{K}_{l_V,m}^{(k)}$, where $l_H$ and $l_V$ represent the numbers of lost $H$- and $V$-polarized photons, respectively. For a single photon, $(l_H,l_V)=(0,0)$, $(1,0)$, and $(0,1)$ reproduce $K_{0,m}^{(k)}$, $K_{1,m}^{(k)}$, and $K_{2,m}^{(k)}$ in \eqref{eq25a}-\eqref{eq25c}, respectively. The reduced states in~\eqref{eq27} and~\eqref{eq28}, together with their multi-photon counterparts, constitute the input to the BSM at $\mathcal{R}_{\rm C}$.
\subsection{BSM and Heralded State}
Following the propagation through the two independent MIMO FSO links, the state of the four photons of the $m$-th attempt before the BSM is $\rho_{A_mC_{1,m}C_{2,m}B_m}^{\mathrm{pre}}\triangleq\rho_{A_mC_{1,m}}^{\Phi^{+}}\otimes
\rho_{C_{2,m}B_m}^{\Phi^{+}}$. Substituting~\eqref{eq27} and~\eqref{eq28}, it is deduced that:
\begin{align}
&\rho_{A_mC_{1,m}C_{2,m}B_m}^{\mathrm{pre}}
= \tau_{1,m}\tau_{2,m}\,\rho^{\Phi^+}_{A_mC_{1,m}}\otimes \rho^{\Phi^+}_{C_{2,m}B_m}
\nn\\
&+\,\tau_{1,m}\left(1-\tau_{2,m}\right)\,
\rho^{\Phi^+}_{A_mC_{1,m}}\otimes\ket{\mathrm{vac}}\bra{\mathrm{vac}}_{C_{2,m}}
\otimes \frac{\mathbf{I}_{B_m}}{2}
\nn\\
&+\,\left(1-\tau_{1,m}\right)\tau_{2,m}\,
\frac{\mathbf{I}_{A_m}}{2}\otimes\ket{\mathrm{vac}}\bra{\mathrm{vac}}_{C_{1,m}}
\otimes\rho^{\Phi^+}_{C_{2,m}B_m}
\nn\\
&+\,\left(1-\tau_{1,m}\right)\left(1-\tau_{2,m}\right)\,
\frac{\mathbf{I}_{A_m}}{2}\otimes\ket{\mathrm{vac}}\bra{\mathrm{vac}}_{C_{1,m}}  \otimes
\nn \\
&\ket{\mathrm{vac}}\bra{\mathrm{vac}}_{C_{2,m}}\otimes\frac{\mathbf{I}_{B_m}}{2}, \quad m= 1,\ldots, r_{\rm ES},
\label{eq30} 
\end{align}
where $\rho^{\Phi^+}$ is given in~\eqref{eq18}, and the four terms correspond to both $C_{1,m}$ and $C_{2,m}$, only $C_{1,m}$, only $C_{2,m}$, and neither of the traveling photons reaching $\mathcal{R}_{\rm C}$, respectively. At $\mathcal{R}_{\rm C}$, a BSM is performed on $C_{1,m}$ and $C_{2,m}$, whose outcomes $\mathcal{M}\in\{\Phi^+,\Phi^-,\Psi^+,\Psi^-\}$ correspond to the projectors:
\beq
\Pi_{\mathcal{M},m} \triangleq \ket{\mathcal{M}}\bra{\mathcal{M}}_{C_{1,m}C_{2,m}},
\label{eq31}
\eeq
where the polarization Bell states are given as follows:
\beq
\ket{\Phi^\pm} \triangleq \frac{\ket{HH}\pm\ket{VV}}{\sqrt{2}},\qquad
\ket{\Psi^\pm} \triangleq \frac{\ket{HV}\pm\ket{VH}}{\sqrt{2}}.
\label{eq32}
\eeq
These projectors act only on the two-photon subspace and therefore have zero support on the last three terms of \eqref{eq30}, in which at least one relay input is vacuum. Since a linear-optics Bell analyzer can unambiguously identify only $\ket{\Psi^\pm}$~\cite{calsamiglia2001}, the accepted outcomes are $\tilde{\mathcal{M}}\in\{\Psi^+,\Psi^-\}$, and all other events are discarded. The probability of an accepted outcome $\tilde{\mathcal{M}}$ on the $m$-th paired eigenchannel is, therefore, given as follows:
\begin{align}
p_{\tilde{\mathcal{M}},m}^{\mathrm{ES}} &\triangleq
\operatorname{Tr}\!\left[\left(\mathbf{I}_{A_m}\otimes\Pi_{\tilde{\mathcal{M}},m}
\otimes\mathbf{I}_{B_m}\right)\rho_{A_mC_{1,m}C_{2,m}B_m}^{\mathrm{pre}}\right]
\nn \\
&= \frac{\tau_{1,m}\tau_{2,m}}{4},
\label{eq33}
\end{align}
and the state shared by Alice and Bob conditioned on this outcome is given by:
\begin{align}
\rho_{A_mB_m|\tilde{\mathcal{M}}}^{\mathrm{ES}}
\triangleq& \frac{1}{p_{\tilde{\mathcal{M}},m}^{\mathrm{ES}}}
\operatorname{Tr}_{C_{1,m}C_{2,m}}\!\Big[
\left(\mathbf{I}_{A_m}\otimes\Pi_{\tilde{\mathcal{M}},m}\otimes\mathbf{I}_{B_m}\right) \nn \\
& \times\rho_{A_mC_{1,m}C_{2,m}B_m}^{\mathrm{pre}}
\left(\mathbf{I}_{A_m}\otimes\Pi_{\tilde{\mathcal{M}},m}\otimes\mathbf{I}_{B_m}\right)\Big] \nn \\
=& \ket{\tilde{\mathcal{M}}}\bra{\tilde{\mathcal{M}}}_{A_mB_m}.
\label{eq34}
\end{align}

An accepted outcome $\Psi^{\pm}$ at $\mathcal{R}_{\rm C}$ projects the retained photons onto the same Bell state $\ket{\Psi^{\pm}}_{A_mB_m}$, although $A_m$ and $B_m$ never interacted. The probability in \eqref{eq33} corresponds to a single accepted Bell outcome on the $m$-th paired eigenchannel. Summing over the two accepted outcomes, the ideal heralding probability is calculated as follows:
\beq
P_{\mathrm{ES},m}^{\mathrm{ideal}} \triangleq
p_{\Psi^+,m}^{\mathrm{ES}} + p_{\Psi^-,m}^{\mathrm{ES}} = \frac{\tau_{1,m}\tau_{2,m}}{2}.
\label{eq35}
\eeq
\section{Performance Analysis}
The previous Section~III characterized entanglement swapping under pure-loss propagation and ideal detection. In practice, the detectors at $\mathcal{R}_{\rm C}$ and at the legitimate terminals have finite efficiencies and produce spurious clicks, so that an accepted BSM announcement may originate either from the heralded state in~\eqref{eq34} or from noise. In addition, the optical setup introduces residual misalignment. In this section, the previously derived ideal heralding probability in~\eqref{eq35} is generalized to the practical yield of accepted
announcements. Then, the heralded state is used to determine the QBER of each origin, and the resulting SKR is derived.
\vspace{-0.2cm}
\subsection{Practical Heralding Yield}
At the relay $\mathcal{R}_{\rm C}$, the BSM is implemented by a linear-optics Bell analyzer with four threshold detectors of efficiency $\eta_R$, in which the outcomes $\Psi^{+}$ and $\Psi^{-}$ are identified by four two-detector click patterns. Each detector produces a spurious click with the following probability:
\beq
p_n \triangleq 1-\left(1-p_d\right) \left(1-p_{\mathrm{bg}}\right),
\label{eq36}
\eeq
where $p_d$ and $p_{\mathrm{bg}}$ represent the independent per-detector probabilities of dark counts and background-induced clicks, respectively. On the $m$-th paired eigenchannel, the probability that both traveling photons produce signal clicks at $\mathcal{R}_{\rm C}$ is given as $m=1,\ldots, r_{ES}$:
\beq
Y_{\mathrm{raw},m} \triangleq T_{1,m}T_{2,m} = \eta_R^{2}\tau_{1,m}\tau_{2,m},
\label{eq37}
\eeq
where $T_{k,m}\triangleq\eta_R\tau_{k,m}$ represents the signal-click probability of the photon from link $k$. An accepted announcement may also be generated when spurious clicks complete an accepted pattern, which yields the noise-induced yield, given as
\beq
Y_{\mathrm{noise},m} \triangleq 2p_n\left(1-p_n\right)^2 P_{1,m} + 4p_n^2\left(1-p_n\right)^2 P_{0,m},
\label{eq38}
\eeq
where $P_{1,m}\triangleq T_{1,m}\left(1-T_{2,m}\right)+T_{2,m}\left(1-T_{1,m}\right)$ and $P_{0,m}\triangleq\left(1-T_{1,m}\right)\left(1-T_{2,m}\right)$ represent the probabilities that exactly one and neither of the photons produce a signal click, respectively, and the coefficients $2p_n\left(1-p_n\right)^2$ and $4p_n^2\left(1-p_n\right)^2$ are the probabilities that one and two spurious clicks, respectively, complete an accepted pattern while the remaining detectors stay silent. Accordingly, the relay gain, defined as the probability that $\mathcal{R}_{\rm C}$ announces an accepted outcome per attempt, is given as follows:
\beq
Q_m \triangleq P_{\mathrm{BSM}}\,Y_{\mathrm{raw},m}\,\left(1-p_n\right)^2 + Y_{\mathrm{noise},m},
\label{eq39}
\eeq
where $P_{\mathrm{BSM}}\triangleq 1/2$ represents the BSM success probability and $\left(1-p_n\right)^2$ denotes the probability that the two remaining detectors produce no spurious click. For ideal detectors at $\eta_R=1$ and $p_n=0$, \eqref{eq39} reduces to $P_{\mathrm{ES},m}^{\mathrm{ideal}}$ in~\eqref{eq35}.
\vspace{-0.2cm}
\subsection{QBER and SKR Analysis}
After the announcement of $\mathcal{R}_{\rm C}$, Alice and Bob retain the attempts with an accepted outcome and perform basis sifting. Since a genuine herald projects the retained photons onto $\ket{\Psi^{\pm}}$ in~\eqref{eq34}, whereas a noise-induced herald yields uncorrelated outcomes, the QBER of the $m$-th paired eigenchannel in basis $\beta\in\{Z,X\}$ is given as follows:
\beq
E_{\beta,m} \triangleq \frac{e_{\beta}P_{\mathrm{BSM}}Y_{\mathrm{raw},m}\left(1-p_n\right)^2
+\frac{1}{2}Y_{\mathrm{noise},m}}{Q_m},
\label{eq40}
\eeq
where $e_\beta$ represents the intrinsic error probability of basis $\beta$:
\beq
e_{\beta} \triangleq
\begin{cases}
e_{\mathrm{opt}}, & \beta = Z \\
V_{\mathrm{HOM}}\,e_{\mathrm{opt}}
+ \dfrac{1}{2}\left(1-V_{\mathrm{HOM}}\right), & \beta = X
\end{cases},
\label{eq41}
\eeq
where $e_{\mathrm{opt}}$ represents the polarization misalignment error and $V_{\mathrm{HOM}}$ denotes the Hong-Ou-Mandel (HOM) visibility at $\mathcal{R}_{\rm C}$. The $X$-basis error is increased by $1-V_{\mathrm{HOM}}$, since the $X$-basis correlation requires the two photons to interfere at $\mathcal{R}_{\rm C}$, whereas the $Z$-basis correlation, established by polarization sorting, is unaffected by their distinguishability. Accordingly, the effective QBER of the MIMO link is given by:
\beq
\mathrm{QBER}_{\mathrm{MIMO}}^{\beta}\triangleq\frac{\mathbb{E}_{\tau_1,\tau_2}\!\left[\sum_{m=1}^{r_{\rm ES}}Q_m E_{\beta, m}\right]}
{\mathbb{E}_{\tau_1,\tau_2}\!\left[\sum_{m=1}^{r_{\rm ES}}Q_m\right]},\, \beta \in\{Z,X\},
\label{eq42}
\eeq

In the asymptotic regime, with the $Z$ basis used for key generation and the $X$ basis for phase-error estimation, the SKR of the $m$-th paired eigenchannel is given as:
\beq
{\rm SKR}_{{\rm ES},m} \triangleq R_{{\rm sift},m}\max\left\{0,\, r_{{\rm sf},m}^{\rm ES}\right\},
\label{eq43}
\eeq
where $R_{{\rm sift},m}\triangleq p_Z^2\eta_A\eta_B Q_m$ represents the sifted key yield, $r_{{\rm sf},m}^{\rm ES} \triangleq 1 - f_{\rm EC}\,h_2(E_{Z,m}) - h_2(E_{X,m})$ is the secret fraction, $h_2(x)\triangleq-x\log_2 x-(1-x)\log_2(1-x)$ denotes the binary entropy function, and $f_{\rm EC}\geq1$ represents the error-correction inefficiency. The paired eigenchannels undergo independent swapping attempts within each pulse, thereby constituting the spatial multiplexing gain of the considered MIMO architecture. Accordingly, the SKR of the MIMO link, averaged over turbulence and pointing fluctuations, is given as
\beq
{\rm SKR}_{\rm ES}^{\rm MIMO} \triangleq \mathbb{E}_{\tau_1,\tau_2}\!\left[\sum_{m=1}^{r_{\rm ES}}
{\rm SKR}_{{\rm ES},m}\right].
\label{eq44}
\eeq
Since each term in~\eqref{eq43} is the product of the sifted key yield and the secret fraction of the corresponding eigenchannel, the effective secret fraction of the MIMO link, which separates the effect of the error rates from that of the yield, is given as:
\beq
r^{\rm ES}_{\rm eff} \triangleq \frac{\mathbb{E}_{\tau_1,\tau_2}\!\left[\sum_{m=1}^{r_{\rm ES}}
Q_m \max\!\left\{0,\, r_{{\rm sf},m}^{\rm ES}\right\}\right]}
{\mathbb{E}_{\tau_1,\tau_2}\!\left[\sum_{m=1}^{r_{\rm ES}} Q_m\right]}.
\label{eq45}
\eeq
\subsection{The Case of Direct-Link between Alice and Bob}
For comparison with the proposed scheme, a direct entanglement-based link is considered, in which Alice measures photon $A$ locally and sends photon $B$ to Bob over a single MIMO FSO link of length $z$, whose channel matrix $\mathbf{H}_{\mathrm{dir}}\in\mathbb{C}^{N_2\times N_1}$ is obtained from~\eqref{eq1} by replacing $z_k$ with $z$. Its SVD is $\mathbf{H}_{\mathrm{dir}} = \mathbf{U}_{\mathrm{dir}}\boldsymbol\Sigma_{\mathrm{dir}}\mathbf{V}_{\mathrm{dir}}^{\dagger}$, whose nonzero singular values $\sigma_{\mathrm{dir},\tilde{m}}$, $\tilde{m}=1,\ldots,r_{\mathrm{dir}}$,
with $r_{\mathrm{dir}}\triangleq{\rm rank}(\mathbf{H}_{\mathrm{dir}})\le\min(N_1,N_2)$, yield the eigenchannel transmissivities $\tau_{\mathrm{dir},\tilde{m}} = \sigma_{\mathrm{dir},\tilde{m}}^{2}$. The gain of the $\tilde{m}$-th direct eigenchannel, defined as the probability of an accepted Alice-Bob coincidence per generated pair, is given as follows:
\beq
Q_{\tilde{m}}^{\mathrm{dir}} \triangleq Y_{\mathrm{sig},\tilde{m}}^{\mathrm{dir}} + Y_{\mathrm{noise},\tilde{m}}^{\mathrm{dir}},
\label{eq47}
\eeq
where $Y^{\rm dir}_{{\rm sig},\tilde{m}}\triangleq\eta_A T_{\mathrm{dir},\tilde{m}} \left(1-p_d\right)\left(1-p_n\right)$ represents the signal-induced coincidence probability, with $T_{\mathrm{dir},\tilde{m}}=\triangleq\eta_B\tau_{\mathrm{dir},\tilde{m}}$, and the noise-induced coincidence probability is:
\begin{align}
&Y_{\mathrm{noise},\tilde{m}}^{\mathrm{dir}} \triangleq \left(1-p_d\right) \left(1-p_n\right)
\Big( 4p_d\,p_n\left(1-\eta_A\right)\left(1-T_{\mathrm{dir},\tilde{m}}\right) 
\nn \\
&\qquad \quad+ 2p_n\,\eta_A\left(1-T_{\mathrm{dir},\tilde{m}}\right) + 2p_d\left(1-\eta_A\right)T_{\mathrm{dir},\tilde{m}} \Big).
\label{eq48}
\end{align}
Since the direct link involves no two-photon interference, its QBER is identical in both bases and given as:
\beq
E_{\tilde{m}}^{\mathrm{dir}}\triangleq \frac{e_{\mathrm{opt}}Y_{\mathrm{sig},\tilde{m}}^{\mathrm{dir}}
+\frac{1}{2}Y_{\mathrm{noise},\tilde{m}}^{\mathrm{dir}}}{Q_{\tilde{m}}^{\mathrm{dir}}},
\label{eq49}
\eeq
and the SKR of the direct MIMO link is given as
\beq
\mathrm{SKR}_{\mathrm{MIMO}}^{\mathrm{dir}} = \mathbb{E}_{\tau_{\mathrm{dir}}}\!\left[
\sum_{\tilde{m}=1}^{r_{\mathrm{dir}}} p_Z^2 Q_{\tilde{m}}^{\mathrm{dir}}
\max\left\{0,\,r_{\rm sf,\tilde{m}}^{\rm dir}\right\}\right],
\label{eq50}
\eeq
where $r_{\rm sf,\tilde m}^{\rm dir}\triangleq 1-(1+f_{\rm EC})\,h_2\big(E^{\mathrm{dir}}_{\tilde m}\big)$ represents the secret fraction of the $\tilde{m}$-th direct eigenchannel. The effective QBER and secret fraction of the direct link, $\mathrm{QBER}_{\mathrm{MIMO}}^{\mathrm{dir}}$ and $r_{\rm eff}^{\rm dir}$, follow from~\eqref{eq42} and~\eqref{eq45} by replacing $Q_m$, $E_{\beta,m}$, and $r_{{\rm sf},m}^{\rm ES}$ with $Q_{\tilde{m}}^{\mathrm{dir}}$, $E_{\tilde{m}}^{\mathrm{dir}}$, and $r_{{\rm sf},\tilde{m}}^{\rm dir}$, respectively.
\section{Numerical Results and Discussion}
This section evaluates the analytical framework presented in Sections~III and~IV using the parameters in Table~\ref{tab1}, which are employed unless stated otherwise. The untrusted relay $\mathcal{R}_{\rm C}$ is located at the midpoint between Alice and Bob, i.e., $z_1=z_2=z/2$, except in Fig.~\ref{f6}, and identical detectors are assumed at the legitimate nodes, i.e., $\eta_A=\eta_B$. Unless stated otherwise, the ideal Bell-pair source is considered, whereas the SPDC and general sources are examined in Section~\ref{sec:spdc}.

\begin{table}[!t]
\caption{The considered simulation parameters.}
\label{tab1}
\centering
\footnotesize
\renewcommand{\arraystretch}{1.1}
\begin{tabular}{@{}p{0.46\columnwidth}p{0.45\columnwidth}@{}}
\toprule
\textbf{Parameter} & \textbf{Value}\\
\midrule
\multicolumn{2}{@{}l}{\underline{\emph{FSO channel and arrays}}}\\
Wavelength, $\lambda$ & 1550 nm\\
Beam waist radius, $w_0$ & 25 mm\\
Receiver aperture radius, $a_r$ & 5 cm\\
Acceptance mode radius, $w_{\rm rx}$ & $\min\{0.89\,a_r,\,w(z_k)\}$ \!\cite{Klein1974OpticalAntennaGain,Ruilier2001}\\
Aperture spacing, $d$ & 12 cm\\
Relay position, & $z_1=z_2=z/2$\\
Absorption, $\delta$ & 0.43 dB/km \cite{sushil_11129674}\\
Structure constant, $C_n^2$ & $10^{-17}$ m$^{-2/3}$ \cite{Andrews2005}\\
Pointing jitter, $\theta_p$ & 1 $\mu$rad (5, 10 $\mu$rad)\\
\midrule
\multicolumn{2}{@{}l}{\underline{\emph{Detectors}}}\\
Detector efficiencies, $\eta_R$, $\eta_A=\eta_B$ & 0.8 (0.2) \cite{marsili2013detecting,Zhang2015InGaAs}\\
Dark count probability, $p_d$ & $10^{-7}$ ($1.6\times10^{-5}$) \cite{marsili2013detecting,Zhang2015InGaAs}\\
Background probability $p_{\rm bg}$ & $10^{-6}$ (night), $10^{-4}$ (day) \cite{Miao2005BGNoise}\\
\midrule
\multicolumn{2}{@{}l}{\underline{\emph{QKD protocol}}}\\
Optical misalignment error, $e_{\rm opt}$ & 0.015 \cite{lo2012mdi}\\
HOM visibility, $V_{\rm HOM}$ & 0.9 (0.8, 0.98)\\
BSM success probability, $P_{\rm BSM}$ & 1/2 \cite{calsamiglia2001}\\
Error-correction inefficiency $f_{\rm EC}$ & 1.1 \cite{Elkouss2010}\\
$Z$-basis probability, $p_Z$ & 0.5\\
Source repetition rate, $f_s$ & 100 MHz\\
\midrule
\multicolumn{2}{@{}l}{\underline{\emph{Sources}}}\\
SPDC mean pair number, $\mu$ & optimized at each $z$ \cite{ma_jul2005}\\
General source, $p_1$, $g$ & $10^{-3}$-$1$, $10^{-3}$-$10$\\
SKR thresholds, ${\rm SKR}_{\rm th}$ & 1, 0.01 bit/s\\
\bottomrule
\end{tabular}
\end{table}
\subsection{Performance with an Ideal Bell-Pair Source}
\begin{figure}[!t]
    \centering
    \includegraphics[width=9cm, height=6.5cm]{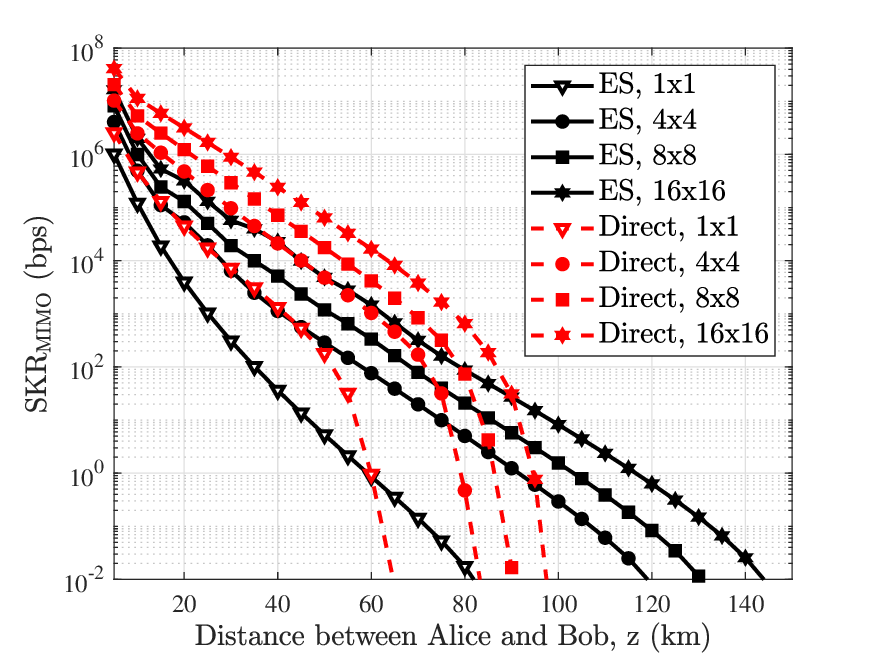}
    \caption{$\mathrm{SKR}_{\rm MIMO}$ versus the distance between Alice and Bob for the entanglement swapping and direct links under a SISO and $4\times4$, $8 \times 8$, and $16\times16$ MIMO configurations at $e_{\rm opt}=0.015$, $p_{\rm bg}=10^{-6}$, and $V_{\rm HOM}=0.9$.}
    \vspace{-0.2cm}
    \label{f2}
\end{figure}
Figure~\ref{f2} illustrates the variation of the SKR of the entanglement swapping and direct link versus the distance between Alice and Bob for the $N=\{1,4,8,16\}$ MIMO configurations. It is observed that the SKR of both architectures decreases with distance, which is attributed to the combined effects of path loss, turbulence-induced fading, and pointing errors on the received optical power. Increasing the MIMO configurations raises the SKR and extends the distance over which a positive key is obtained. 
More importantly, the entanglement-swapping link sustains a positive SKR well beyond the distance at which the direct link collapses. This advantage is attributed to the coincidence requirement at the relay, under which noise competes with the transmissivity of each half-length segment rather than that of the full Alice-Bob path, making the entanglement-swapping link considerably more robust to noise and channel loss. It is also observed that at short distances, the direct link achieves a higher SKR, so that the SKR curves of the two links intersect. The distance at which this occurs, referred to as the crossover distance, is given as
\beq
z_{\rm c} = \min\left\{ z : {\rm SKR}_{\rm ES}^{\rm MIMO}(z) \ge
{\rm SKR}_{\rm dir}^{\rm MIMO}(z) \right\},
\label{eq51}
\eeq
where ${\rm SKR}_{\rm ES}^{\rm MIMO}$ and ${\rm SKR}_{\rm dir}^{\rm MIMO}$ are given in~\eqref{eq44} and \eqref{eq50}, respectively, and $z_{\rm c}$ represents the distance beyond which the entanglement swapping link provides the higher SKR. Notably, $z_{\rm c}$ increases with the MIMO configuration, and the reach advantage of entanglement swapping is preserved for all considered configurations, confirming it as the preferred architecture for long-distance FSO-QKD.

\begin{figure}[!t]
    \centering
    \includegraphics[width=9cm, height=6.5cm]{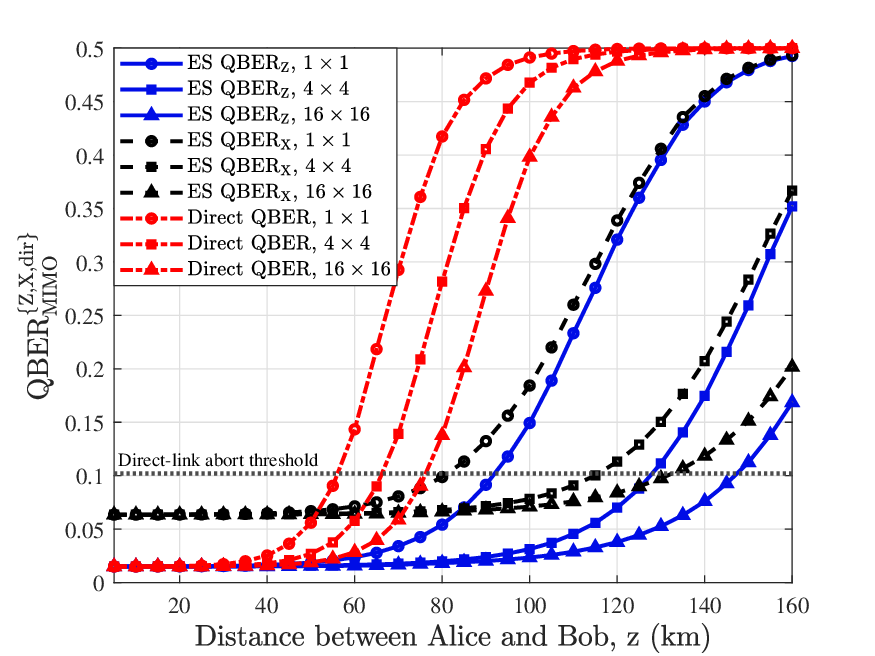}
    \caption{$\mathrm{QBER}_{\rm MIMO}$ versus the distance between Alice and Bob for the entanglement swapping and direct links for $N=\{1,4,16\}$ MIMO configurations at $e_{\rm opt}=0.015$, $p_{\rm bg}=10^{-6}$, and $V_{\rm HOM}=0.9$.}
    \vspace{-0.2cm}
    \label{f3}
\end{figure}
Figure~\ref{f3} illustrates the $Z$- and $X$-basis QBERs of the entanglement swapping link and the direct-link QBER versus distance between Alice and Bob for $N=\{1,4,16\}$ MIMO configurations. It is observed that all QBERs remain nearly constant at short distances and gradually increase towards $0.5$ as the distance increases. This is attributed to a decrease in received optical power due to path loss and turbulence-induced fading, allowing background and dark counts to dominate the detection events. We can notice that, at short distances, $\mathrm{QBER}^{Z}_{\mathrm{MIMO}}$ and $\mathrm{QBER}^{\mathrm{dir}}_{\mathrm{MIMO}}$ approach the optical misalignment error, whereas $\mathrm{QBER}^{X}_{\mathrm{MIMO}}$ settles at the higher value. This is because the $Z$-basis correlation relies only on polarization sorting at the relay $\mathcal{R}_{\rm C}$, while the $X$-basis correlation requires coherent two-photon interference, so that imperfect interference visibility introduces an additional error that is absent in the direct link. It is also observed that the entanglement swapping link maintains a QBER below the abort threshold over a significantly longer distance than the direct link, since each half-link spans only half the Alice-Bob distance and hence experiences considerably lower loss. As the MIMO configuration increases, the QBER curves shift towards larger distances, which is attributed to the enhanced spatial diversity and the averaging of turbulence-induced fading across multiple optical paths. However, it is important to note that the MIMO configuration does not affect QBER at short distances, as this value is determined by the transceiver optics rather than the channel.

\begin{figure}[!t]
    \centering
    \includegraphics[width=9cm, height=6.5cm]{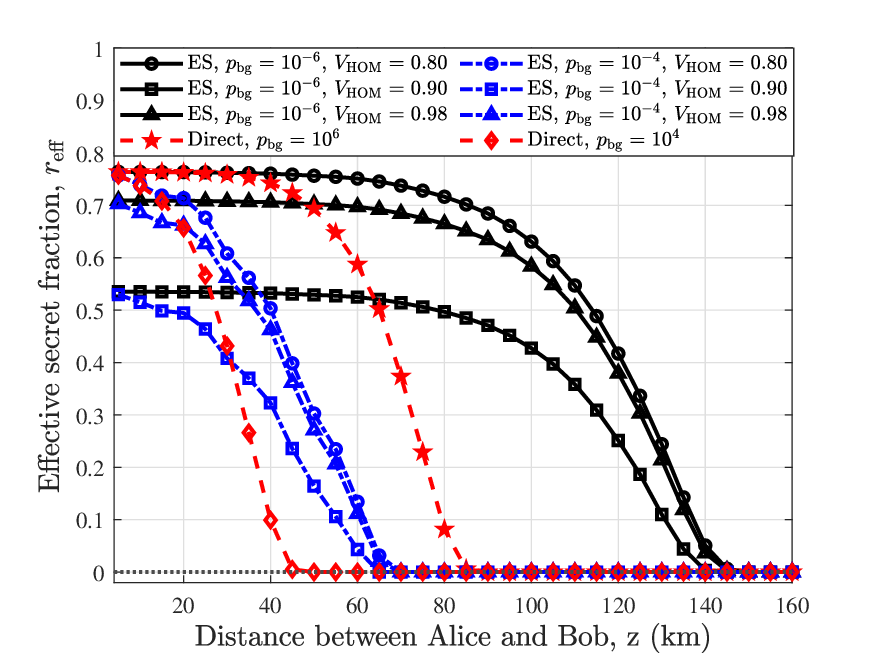}
    \caption{$ r_{\rm eff}$ versus distance between Alice and Bob for the entanglement swapping link with $V_{\rm HOM} = \{0.80, 0.90, 0.98\}$, and for the direct link, for the $8\times8$ MIMO at $p_{\rm bg}=\{10^{-6}, 10^{-4}\}$.}
    \vspace{-0.2cm}
    \label{f4}
\end{figure}
Figure~\ref{f4} illustrates the effective secret fraction versus the distance between Alice and Bob for the entanglement swapping link with $V_{\rm HOM} = \{0.80, 0.90, 0.98\}$, and the direct link under night ($p_{\rm bg}=10^{-6}$) and day ($p_{\rm bg}=10^{-4}$) conditions for $8\times8$ MIMO configuration. It is observed that, at short distances, $r_{\mathrm{eff}}$ of the entanglement swapping link remains nearly constant for each $V_{\rm HOM}$, while the direct link performs better.
This difference is attributed to the imperfect two-photon interference, which increases only the $X$-basis error rate, leaving the $Z$-basis error rate and direct link unaffected. At longer distances, background and dark counts cause an abrupt decline in the direct link, whereas entanglement swapping exhibits a gradual decay and maintains a positive secret fraction over longer distances at night. The entanglement swapping link surpasses the direct link beyond a crossover distance that decreases with increasing $V_{\rm HOM}$. Higher daytime background noise reduces the operating range of both links, despite the noise-filtering advantage of entanglement swapping.

\begin{figure}[!t]
    \centering
    \includegraphics[width=9.5cm, height=7cm]{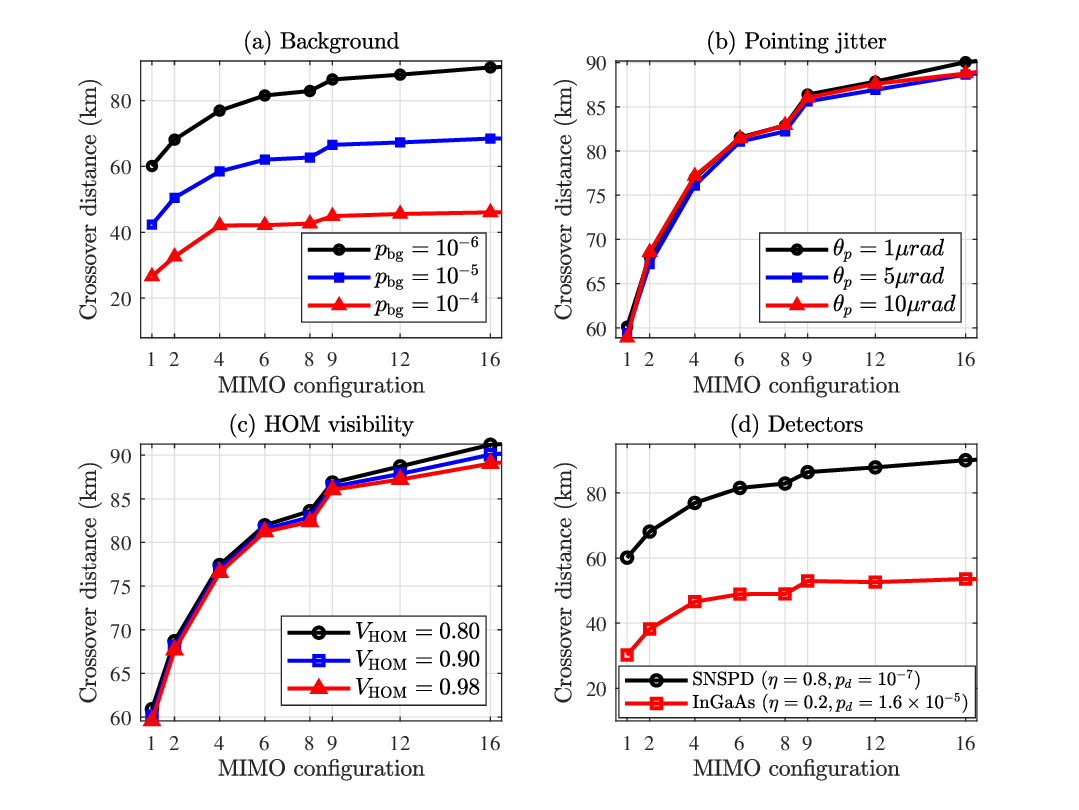}
    \caption{Crossover distance versus the MIMO configuration $N$ for different: (a) background, (b) pointing jitter, (c) HOM visibility, and (d) detector parameters.}
    \vspace{-0.2cm}
    \label{f5}
\end{figure}
Figure~\ref{f5} illustrates the crossover distance $z_{\rm c}$ in \eqref{eq51} versus the MIMO configuration for the ideal Bell-pair source under different (a) background click probabilities, (b) pointing jitters, (c) HOM visibilities, and (d) detector technologies, with all remaining parameters set to their baseline values in Table~\ref{tab1} under night conditions. It is observed that $z_{\rm c}$ increases with the MIMO configuration in all cases and gradually saturates for larger MIMO, which is attributed to the additional eigenchannels that raise the SKR of the entanglement-swapping link near the noise-limited cutoff of the direct link, whereas the saturation arises because the weaker eigenchannels contribute progressively less to the SKR. Moreover, $z_{\rm c}$ is considerably reduced by a stronger background and by InGaAs detectors, since their higher noise moves the cutoff of the direct link to shorter distances, whereas it is nearly insensitive to $\theta_p$ and $V_{\rm HOM}$, which mainly reduce the SKR at $z_{\rm c}$ rather than shift it. Notably, the noise environment and detector technology therefore determine where entanglement swapping becomes beneficial, whereas MIMO improves both links and shifts this point to
longer distances.
\subsection{Impact of the Relay Position}
The preceding results assume that the relay is located at the midpoint between Alice and Bob. In practice, however, the relay location is often constrained by the terrain and the available infrastructure. Moreover, while $z_{\rm c}$ identifies where entanglement swapping outperforms the direct link, the maximum distance over which each link sustains a given key rate is equally relevant. For a target rate ${\rm SKR}_{\rm th}$, the reach of a link is defined as follows:
\beq
z_{\rm th} \triangleq \max\left\{ z : {\rm SKR}^{\rm MIMO}(z) \ge {\rm SKR}_{\rm th} \right\},
\label{eq52}
\eeq
and the reach gain of entanglement swapping is given as
\beq
\Delta z \triangleq z_{\rm th}^{\rm ES} - z_{\rm th}^{\rm dir},
\label{eq53}
\eeq
where $z_{\rm th}^{\rm ES}$ and $z_{\rm th}^{\rm dir}$ represent the reaches of the entanglement-swapping and direct links, respectively, both employing the same source, so that a positive $\Delta z$ indicates that the relay extends the distance over which a key rate of at least ${\rm SKR}_{\rm th}$ is obtained.

\begin{figure}[!t]
\centering
\includegraphics[width=\columnwidth]{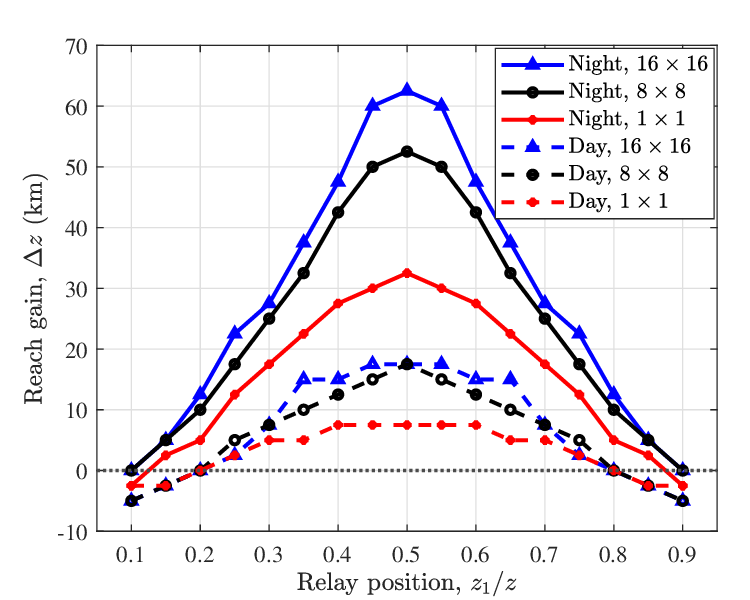}
\caption{Reach gain versus the normalized relay position for the ideal Bell-pair source under night and day conditions.}
\vspace{-0.2cm}
\label{f6}
\end{figure}
Figure~\ref{f6} illustrates $\Delta z$ in~\eqref{eq53} versus the normalized relay position $z_1/z$ for the ideal Bell-pair source for $N=\{1,8,16\}$ MIMO configurations under night and day conditions, where $z_2 = z - z_1$ and the reach is evaluated for a positive SKR, i.e., ${\rm SKR}_{\rm th} \to 0$. It is shown that $\Delta z$ is maximized when the relay is located at the midpoint and decreases as the relay moves toward either terminal. This is attributed to the product $\tau_{1,m}\tau_{2,m}$ in~\eqref{eq37}, since an accepted event requires both photons to reach $\mathcal{R}_{\rm C}$, so that the longer segment increasingly limits the performance, and the benefit of dividing the link is gradually lost. Nevertheless, $\Delta z$ remains positive over a broad range of relay positions, implying that the relay need not be placed exactly at the midpoint, which eases practical deployment. This range is narrower under day conditions, where the stronger background shortens the reach of both links. Notably, a higher MIMO configuration increases reach gain but doesn’t significantly extend range, implying that MIMO enhances the performance of optimally placed relays rather than compensating for suboptimal positions.
\subsection{Impact of a Practical SPDC Source}
\label{sec:spdc}
To assess whether the advantages obtained with the ideal Bell-pair source persist with a practical source, the SPDC source in~\eqref{eq15} is considered at both Alice and Bob for the entanglement-swapping link, and at Alice for the direct link. For the entanglement-swapping link, the yields and error rates are evaluated numerically by applying~\eqref{eq29} to the multi-photon states in~\eqref{eq14}, retaining up to three pairs per source, which is sufficient since the optimal $\mu$ remains well below unity. For the direct link, the closed-form gain and QBER of a thermal entangled-photon source in~\cite{ma_jul2005} are used, which follow from the same pure-loss channel and threshold-detector model. Since increasing $\mu$ raises both the single-pair and the multi-pair emission probabilities, $\mu$ is optimized separately for each link at each distance, and the QBER is evaluated at the optimal $\mu$.

\begin{figure}[!t]
    \centering
    \includegraphics[width=9cm, height=8cm]{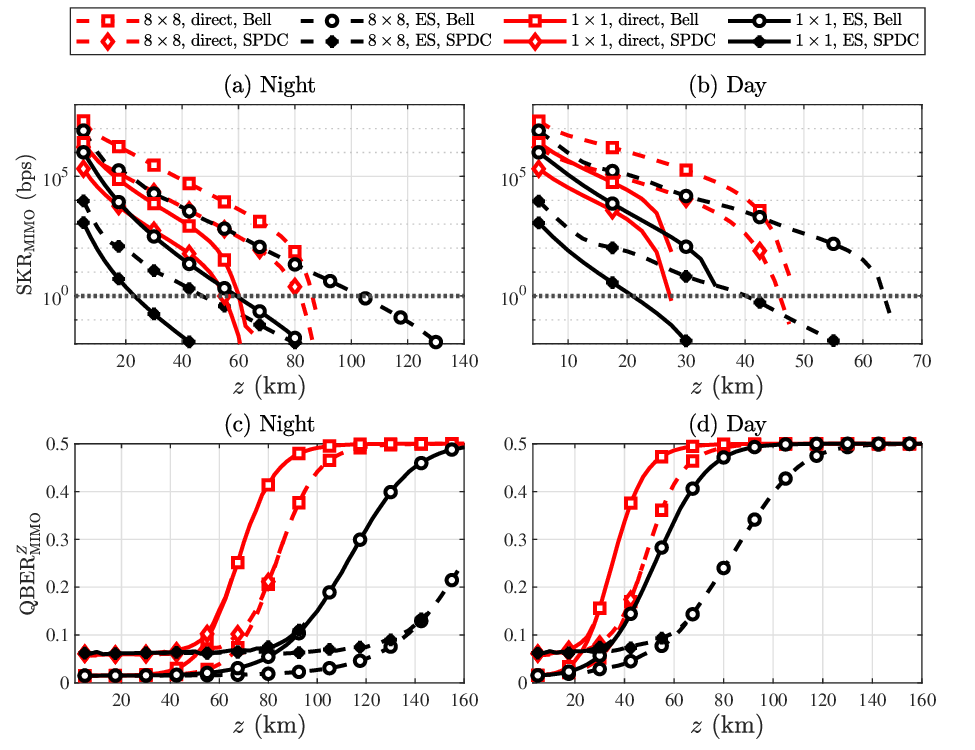}
    \caption{SKR and QBER versus distance for the ideal Bell-pair and SPDC sources with and without entanglement swapping under (a), (c) night and (b), (d) day conditions.}
    \vspace{-0.2cm}
    \label{f7}
\end{figure}
\begin{figure*}[!t]
\centering
\includegraphics[width=\textwidth]{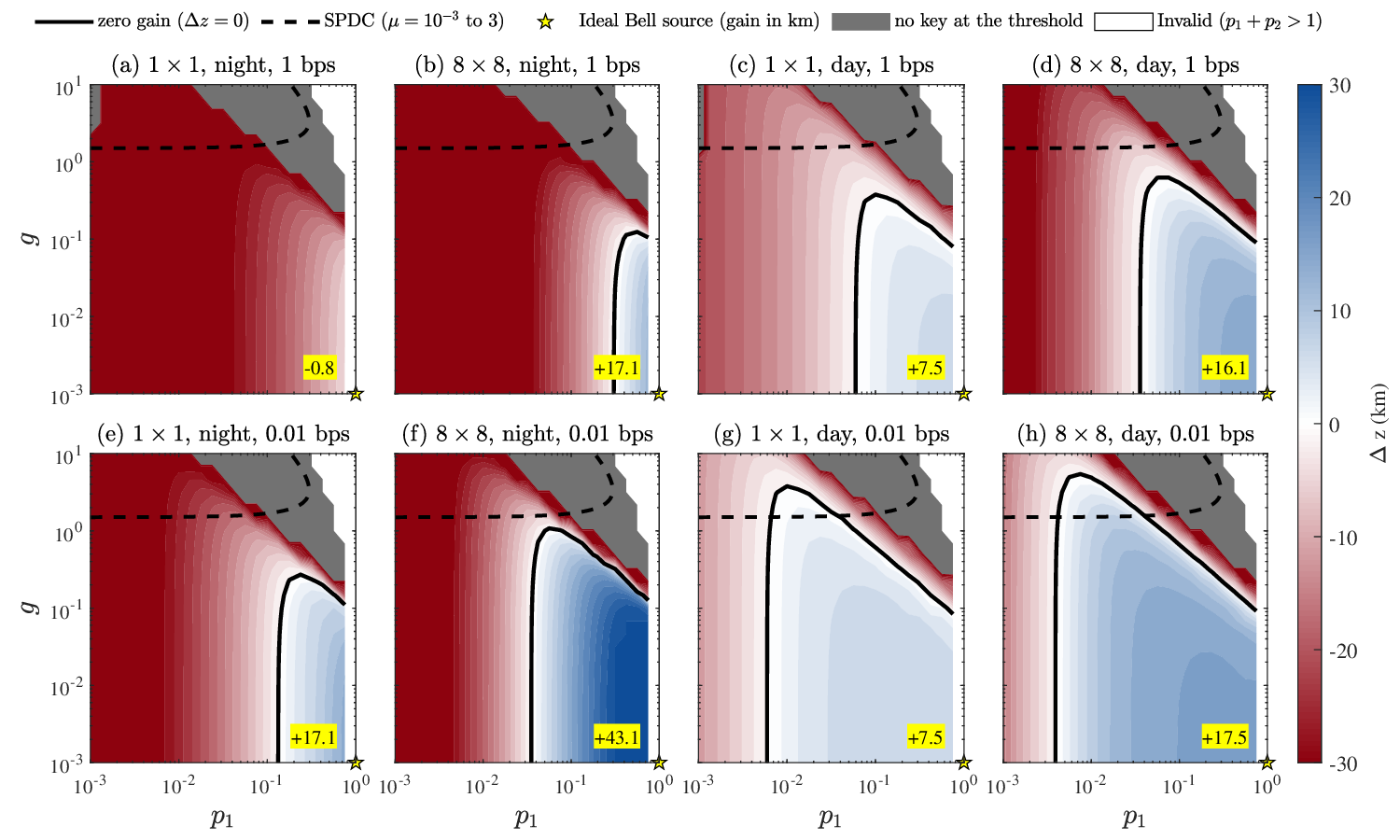}
\caption{Reach gain of entanglement swapping over the direct link versus the single-pair probability $p_1$ and the multi-pair parameter $g$ of the source, considering SKR thresholds of (a)-(d) $1$~bit/s and (e)-(h) $0.01$~bit/s.}
\vspace{-0.2cm}
\label{f8}
\end{figure*}
Figure~\ref{f7} illustrates the SKR and the $Z$-basis QBER versus the distance between Alice and Bob for the entanglement-swapping and direct links with the ideal Bell-pair and SPDC sources, for $N=\{1,8\}$ MIMO configurations under night and day conditions. It is observed in Fig.~\ref{f7}(a) and (b) that, with the ideal Bell-pair source, the entanglement-swapping link sustains a positive SKR beyond the cutoff of the direct link in all cases, and this advantage is more pronounced under day conditions and for the higher MIMO configuration. With the SPDC source, the SKR of the direct link decreases only moderately and collapses at the same distance, whereas that of the entanglement-swapping link decreases substantially and remains below the direct link over most of the distance range. This is attributed to the fact that an accepted swapping event requires a single-pair emission from both sources, so that its yield scales with the square of the single-pair probability. Moreover, MIMO raises the SKR of both SPDC links by a comparable factor, thereby extending their reach but not closing the gap between them. Fig.~\ref{f7}(c) and (d) show that both SPDC links exhibit a higher QBER floor than the ideal source, since the optimal $\mu$ balances the single-pair yield against multi-pair emissions. Notably, for both sources, the QBER of the entanglement-swapping link rises at considerably longer distances than that of the direct link, implying that the noise-filtering effect of the coincidence requirement at $\mathcal{R}_{\rm C}$ is retained with SPDC, and that the SPDC penalty arises mainly from the reduced yield rather than from the error rate.

To identify the source quality for which entanglement swapping extends the reach, Fig.~\ref{f8} illustrates $\Delta z$ in~\eqref{eq53} versus $p_1$ and $g$ of the general source in~\eqref{eq16} for $N=\{1,8\}$ MIMO configurations under night and day conditions and for ${\rm SKR}_{\rm th}=\{0.01, 1\}$ bps, where the solid black line marks $\Delta z = 0$ and the dashed black line represents the SPDC source. It is observed that a positive reach gain requires a sufficiently large $p_1$ together with a sufficiently small $g$. The lower bound on $p_1$ is attributed to the yield of the entanglement-swapping link, which scales with $p_1^2$ rather than $p_1$, so that a weak source drives its SKR below ${\rm SKR}_{\rm th}$ before the direct link collapses. Consistently, lowering ${\rm SKR}_{\rm th}$ relaxes this bound. The upper bound on $g$ is attributed to multi-pair emissions, and it tightens as $p_1$ increases, since $p_2$ in~\eqref{eq16} grows quadratically with $p_1$. Moreover, the region of positive gain is considerably larger under day conditions, where the stronger background shortens the direct link, whereas under night conditions with the higher threshold, the $1\times1$ configuration provides no reach gain even for the ideal source. Notably, the $8\times8$ configuration enlarges the region of positive gain in all cases by lowering the required $p_1$ and raising the tolerable $g$. In contrast, the SPDC source lies outside this region in all cases except under day conditions with the lower threshold, where it provides only a small gain. Overall, these results indicate that entanglement swapping extends the reach of terrestrial MIMO-FSO links only with a near-deterministic entangled-photon source, and that MIMO
considerably relaxes the source-quality requirements to achieve this extension.
\section{Conclusions}
This paper presented an entanglement-swapping-based DV-QKD system over MIMO FSO links, in which two legitimate users establish polarization entanglement via an untrusted relay for secret key exchange. The considered FSO channels were subjected to absorption, diffraction, beam wander, pointing error, log-normal turbulence-induced fading, and correlated path phases, and were assumed to be potentially controlled together with the relay by an eavesdropper. A novel mathematical framework for characterizing the MIMO FSO channel was presented, in which the SVD decouples each segment into spatial eigenchannels, which are then paired at the relay. Furthermore, photon loss was described by a Kraus operator-sum representation, and the post-measurement state after the BSM was derived. Based on a threshold-detector model including dark counts, background noise, and imperfect two-photon interference, closed-form expressions for the gain, QBER, and SKR of the entanglement-swapping link were derived and compared with those of a direct entanglement-based link built on the same models. Numerical investigations showcased that entanglement swapping extends the reach beyond the noise-limited cutoff of the direct link at the cost of a lower SKR at short distances. It was also demonstrated that the crossover distance is determined primarily by the background noise and the detector technology, and increases with the MIMO configuration. The reach gain was found to be maximized for a midpoint relay, remaining positive over a broad range of relay positions. Moreover, for practical sources, the reach extension was shown to require a high single-pair and a low multi-pair emission probability, which an SPDC source does not satisfy, whereas increasing the MIMO configuration substantially relaxed this requirement. 

Future work includes extending the proposed framework to finite-key security analysis, moderate-to-strong turbulence, and imperfect channel state information in the eigenchannel processing. Moreover, multiplexed, heralded, and quantum-dot entangled-photon sources will be considered, and entanglement-swapping-based DV-QKD architecture will be extended to multi-user and multi-hop MIMO FSO networks.
\bibliographystyle{IEEEtran}
\bibliography{IEEEabrv,bibliography}
\end{document}